\documentclass[12pt]{article}
\usepackage{layout,amssymb,amsmath,epsfig}

\begin{document}
\title{\bf Energy conditions in $f(T)$ gravity with non-minimal torsion-matter coupling}
\author{M. Zubair \thanks{mzubairkk@gmail.com; drmzubair@ciitlahore.edu.pk} and Saira Waheed
\thanks{smathematics@hotmail.com}\\\\
Department of Mathematics, COMSATS\\
Institute of Information Technology Lahore, Pakistan.}

\date{}

\maketitle

\begin{abstract}
The present paper examines the validity of energy bounds in a
modified theory of gravity involving non-minimal coupling of torsion
scalar and perfect fluid matter. In this respect, we formulate the
general inequalities of energy conditions by assuming the flat FRW
universe. For the application of these bounds, we particularly focus
on two specific models that are recently proposed in literature and
also choose the power law cosmology. We find the feasible
constraints on the involved free parameters and evaluate their
possible ranges graphically for the consistency of these energy
bounds.
\end{abstract}

\section{Introduction}

The recent speedy expansion of our cosmos is one of the most
attractive advances on the observational landscape of fundamental
physics. This expanding paradigm has been corroborated by the
observational probes of numerous mounting astronomical evidences
like type Ia supernovae (SNe Ia), the cosmic microwave background
(CMB), large scale structure surveys (LSS) and Wilkinson Microwave
Anisotropy Probe (WMAP) (Riess et al. 1998; Perlmutter 1999; Bennett
2003; Allen et al. 2004; Tegmark et al. 2004; Spergel et al. 2007).
In evolution picture of cosmos, two stages of cosmic acceleration
has been put forward by the researchers, the primordial inflationary
stage (before radiation state of cosmos) and the recent as well as
the final cosmic phases (after the matter dominated state). Inspite
of all fascinating aspects of general relativity (GR), it gives rise
to decelerated phases of cosmos as the nature of gravitational force
is attractive and hence turns out to be incompatible with this
primal fact. Thus it leads to the speculation that there is some
other mysterious anti-gravitational source with some unusual
particulars causing this cosmic expansion faster and is labeled as
dark energy (DE).

The investigation of the complete evolutionary cosmic picture, from
the Big Bang era to its final fate, has been widespread among the
scientists. For this purpose, numerous efforts have been made based
upon different strategies. These efforts are mainly grouped into two
kinds: modifications in the matter sector of the Einstein-Hilbert
Lagrangian density and the extension of the gravitational framework
of GR by introducing some terms representing the DE source. The
cosmological constant (Peebles and Ratra 2003), Chaplygin gas matter
with its different modified versions (Bento et al. 2002; Benaoum
2012), scalar field models like quintessence (Caldwell et al. 1998;
Steinhardt et al. 1999) are some leading examples of the candidates
of the first group, while the modified theories of gravity are the
representatives of the second group (Lobo 2008; Clifton et al.
2012). Some well-motivated examples of such theories include $f(R)$
gravity (the Ricci scalar of the Einstein-Hilbert action functional
is replaced by a generic function $f(R)$) (Nojiri and Odintsov 2011;
Bamba et al. 2012a), Gauss-Bonnet gravity (including Gauss-Bonnet
invariant term) (Cognola et al. 2006; Li et al. 2007), $f(T)$ theory
(based on the torsion tensor as well as its corresponding scalar)
(Ferraro and Fiorini 2007; Bamba et al. 2012b; Setare and Darabi
2012; Setare and Mohammadipour 2012, 2013), $f(R,T)$ gravity (where
$T$ is the trace of the energy-momentum tensor) (Harko et al. 2011;
Sharif and Zubair 2014a, 2014b, 2014c, 2014d) and the scalar-tensor
theories (based on both scalar and tensor fields) (Fujii and Maeda
2004; Brans 2005; Faraoni 2004) etc.

Recently, the introduction of non-minimal coupling between matter
and curvature in the context of modified theories has become a
center of interest for the researchers. Bertolami et al. (2007)
derived the dynamical equation for massive particles in $f(R)$
gravity by assuming an explicit interaction between the scalar
curvature and density of matter. They concluded that the presence of
this coupling yields an extra force. Bertolami and Sequeira (2009)
discussed the energy condition bounds in $f(R)$ gravity involving a
non-minimal interaction of curvature and matter and investigated the
stability via Dolgov-Kawasaki criterion. Bertolami and Páramos
(2014) studied the modification of Friedmann equation due to the
inclusion of non-minimal coupling and provided two ways to handle
the corresponding situation. Furthermore, they addressed the
cosmological constant problem in such theory.

Another important and conceptually rich class consists of
gravitational modifications involving torsion description of
gravity. It is interesting to mention here that teleparallel
equivalent of GR has been constructed by Einstein himself by
including torsionless Levi-Civita connection instead of
curvatureless Weitzenböck connection and the vierbein as the
fundamental ingredient for the theory (Moller 1961; Pellegrini and
Plebanski 1963; Hayashi and Shirafuji 1979; Maluf 2013).
Consequently, the corresponding formulation replaces the Ricci
tensor and Ricci scalar by the torsion tensor and torsion scalar
respectively. Further, its modified form has been proposed and
discussed by numerous authors like (Ferraro and Fiorini 2007;
Bengochea and Ferraro 2009; Bamba and Geng 2011). Harko et al.
(2014) constructed a more general type of $f(T)$ gravity by
introducing a non-minimal interaction of torsion with matter in the
Lagrangian density. They discussed the cosmological implications of
this theory and concluded that the universe model may correspond to
de Sitter, dark-energy-dominated, accelerating phase when model
parameters are assigned to large values.

The energy bounds have been widespread to investigate various issues
in GR and cosmology (Visser 1997; Santos and Alcaniz 2006; Santos et
al. 2006; Gong, Y. et al. 2007; Gong and Wang 2007). Energy bounds
have been explored in different modified theories to constrain the
free variables like scalar-tensor theory (Sharif and Waheed 2013),
modified Gauss-Bonnet gravity (Garcia 2011; Zhao 2012), $f(R)$
gravity (Santos et al. 2007; Santos et al. 2010), $f(T)$ gravity
where $T$ is torsion scalar (Liu and Reboucas 2012), $f(R)$ gravity
with nonminimal coupling to matter (Wang et al. 2010; Bertolami and
Sequeira 2009), $f(R,\mathcal{L}_m)$ gravity (Wang and Liao 2012),
$f(R,T)$ gravity (Sharif and Zubair 2013a) and
$f(R,T,R_{\mu\nu}T^{\mu\nu})$ gravity (Sharif and Zubair 2013b).
Sharif and Waheed (2013) explored the energy condition bounds in the
most general scalar-tensor theory involving second-order derivatives
of scalar field and then discuss these conditions for different
cases. Sharif and Zubair (2013a) have investigated the energy bounds
in $f(R,T)$ gravity and by selecting a particular class of models,
they studied the stability of power law solutions. In another paper
(2013b), the same authors discussed the validity of the energy
bounds in a newly modified gravity labeled as
$f(R,T,R_{\mu\nu}T^{\mu\nu})$ gravity and also explore
Dolgov-Kawasaki instability for two specific $f(R,T)$ models.

In the present work, we deal with the energy conditions bounds in a
modified theory of gravity involving a non-minimal coupling of
torsion scalar and matter by taking flat FRW model filled with
perfect fluid. The paper has been designed in the following outline.
In the next section, we provide the basic formulation of the field
equations of this gravity. Section \textbf{3} provides a brief
description of the energy bounds in the context of GR and also their
extension to modified frameworks of gravity. In the same section, we
analyze the obtained inequalities by choosing two recently proposed
models. Section \textbf{4} concludes the whole discussion.

\section{Modified $f(T)$ gravity with non-minimal torsion-matter coupling}

A more general $f(T)$ gravity involving non-minimal coupling between
torsion scalar and matter Lagrangian is defined by the action (Harko
et al. 2014)
\begin{equation}\label{1}
\mathcal{A}=\frac{1}{2{\kappa}^2}\int{dx^4e\{T+f_1(T)+[1+\lambda{f_2(T)}]
\mathcal{L}_{m}\}},
\end{equation}
where $\kappa^2=8\pi{G}$, $f_i(T)$(i=1,2) are arbitrary functions of
torsion scalar, $\lambda$ is the coupling parameter and
$\mathcal{L}_{m}$ denotes the Lagrangian density of matter part. The
field equations in non-minimal $f(T)$ theory can be determined by
varying the action with respect to the tetrad $e^\mu_i$ as
\begin{eqnarray}\nonumber
&&(1+f'_1+\lambda{f}'_2\mathcal{L}_m)\left[e^{-1}\partial_\mu(ee_i^{\sigma}{S}_\sigma^{\rho\mu})
-e^\sigma_iT^\mu_{\nu\sigma}S^{\nu\rho}_\mu\right]+(f''_1+\lambda{f}''_2\mathcal{L}_m)
\\\nonumber&\times&\partial_\mu{T}e^\sigma_iS_\sigma^{\rho\mu}+\frac{1}{4}e^\rho_i(f_1+T)-\frac{1}{4}
\lambda{f}'_2\partial_\mu{T}e^\sigma_iS_\sigma^{(m)\rho\mu}+\lambda{f}'_2e^\sigma_i
S_\sigma^{\rho\mu}\partial_\mu{\mathcal{L}}_m\\\label{2}&=&4\pi{G}(1+\lambda{f_2})e^\sigma_iT_\sigma^{(m)\rho},
\end{eqnarray}
where prime indicates differentiation with respect to torsion scalar
and $S_i^{(m)\rho\mu}$ is defined as
\begin{equation}\label{3}
S_i^{(m)\rho\mu}=\frac{\partial{\mathcal{L}_m}}{\partial\partial_\mu{e}^i_\rho}.
\end{equation}
Equation (\ref{2}) reduces to the field equations in $f(T)$ theory
of gravity for $\lambda=0$ or $f_2(T)=0$. The contribution to the
energy momentum tensor of matter is defined as
\begin{equation}\label{4}
T_{{\mu}{\nu}}=({\rho}+p)u_{\mu}u_{\nu}-pg_{{\mu}{\nu}},
\end{equation}
where energy density and pressure are denoted by $\rho$ and $p$. We
set the matter Lagrangian density as $L_m=-\rho$, which implies
$S_i^{(m)\rho\mu}=0$. We take the homogeneous and isotropic flat FRW
metric defined as
\begin{equation}\label{5}
ds^{2}=dt^2-a^2(t)d\textbf{x}^2,
\end{equation}
where $a(t)$ represents the scale factor and $d\textbf{x}^2$
contains the spatial part of the metric and corresponding tetrad
components are $e^i_\mu=(1,a(t),a(t),a(t))$. In the FRW background,
the field equations may be written as
\begin{eqnarray}\label{6}
3H^2=8\pi{G}[1+\lambda(f_2+12H^2f'_2)]\rho-\frac{1}{6}(f_1+12H^2f'_1),\\\label{7}
\dot{H}=-\frac{4\pi{G}(\rho+p)[1+\lambda(f_2+12H^2f'_2)]}{1+f'_1-12H^2f''_1-16\pi{G}\lambda\rho
(f'_2-12H^2f''_2)},
\end{eqnarray}
where $H=\dot{a}/a$ is the Hubble parameter and dot denotes the
derivative with respect to cosmic time $t$. Equations (\ref{6}) and
(\ref{7}) can be expressed as
\begin{eqnarray}\label{8}
3H^2=8\pi{G}\rho_{eff},\quad -(2\dot{H}+3H^2)=p_{eff},
\end{eqnarray}
where $\rho_{eff}$ and $p_{eff}$ are the energy density and pressure
respectively, defined by
\begin{eqnarray}\label{9}
\rho_{eff}&=&[1+\lambda(f_2+12H^2f'_2)]\rho-\frac{1}{16\pi{G}}(f_1+12H^2f'_1),\\\label{10}
p_{eff}&=&(\rho+p)\frac{[1+\lambda(f_2+12H^2f'_2)]}{1+f'_1-12H^2f''_1-16\pi{G}\lambda\rho
(f'_2-12H^2f''_2)}\\\nonumber&+&\frac{1}{16\pi{G}}(f_1+12H^2f'_1)-[1+\lambda(f_2+12H^2f'_2)]\rho.
\end{eqnarray}

\section{Energy Conditions}

Here firstly, we will discuss the general procedure for energy
condition bounds in GR and then extend it to modified gravity
theories. Then by following the outlined procedure, we concentrate
on flat FRW model filled with perfect fluid matter. Further we
investigate these bounds by focusing on two different models of
$f(T)$ gravity.

\subsection{Raychaudhuri Equation}

In order to comprehend various cosmological geometries and some
general results associated with the strong gravitational fields,
energy bounds are of great interest. In GR, there are four explicit
forms of energy conditions namely: the strong (SEC), null (NEC),
dominant (DEC) and weak energy conditions (WEC) (Hawking and Ellis
1973; Carroll 2004). Basically, the SEC and NEC arise from the
fundamental characteristic of gravitational force that it is
attractive along with a well-known geometrical result describing the
dynamics of nearby matter bits known as Raychaudhari equation. The
Raychaudhari equation specifies the temporal evolution of expansion
scalar $\theta$ in terms of some tensorial quantities like Ricci
tensor, shear tensor $\sigma^{\mu\nu}$ and rotation
$\omega^{\mu\nu}$ for both the time and lightlike curves.
Mathematically, it is given by relations
\begin{eqnarray}\label{12*}
\frac{d\theta}{d\tau}=-\frac{1}{3}\theta^2-\sigma_{\mu\nu}\sigma^{\mu\nu}
+\omega_{\mu\nu}\omega^{\mu\nu}-R_{\mu\nu}u^\mu u^\nu,\\\label{13*}
\frac{d\theta}{d\tau}=-\frac{1}{3}\theta^2-\sigma_{\mu\nu}\sigma^{\mu\nu}
+\omega_{\mu\nu}\omega^{\mu\nu}-R_{\mu\nu}k^\mu k^\nu.
\end{eqnarray}

Due to attractive nature of gravity, geodesics become closer to each
other satisfying $\frac{d\theta}{d\tau}<0$ and hence yields
converging time and lightlike congruences. To simplify the resulting
inequalities, we can ignore the quadratic terms by taking the
assumptions that there are infinitesimal distortions in geodesics
(time or null) which is hypersurface orthogonal as well, i.e.,
$\omega_{\mu\nu}=0$ (no rotation). Consequently, integration of the
simplified Raychaudhari equations leads to $\theta=-\tau
R_{\mu\nu}u^\mu u^\nu=-\tau R_{\mu\nu}k^\mu k^\nu$ for timelike and
null geodesics, respectively. Using $\frac{d\theta}{d\tau}<0$, this
can also be rearranged to
\begin{equation*}
R_{\mu\nu}u^\mu u^\nu\geq0,\quad R_{\mu\nu}k^\mu k^\nu\geq0.
\end{equation*}

Since a relation of Ricci tensor in terms of energy-momentum tensor
and its trace can be found by inverting the gravitational field
equations (which interrelates both curvature (Ricci tensor) and
matter sectors) as follows
\begin{eqnarray}\label{14**}
R_{\mu\nu}=T_{\mu\nu}-\frac{T}{2}g_{\mu\nu}.
\end{eqnarray}
Therefore the inequalities of energy bounds take the following forms
\begin{eqnarray}\label{14*}
&&R_{\mu\nu}u^\mu u^\nu=(T_{\mu\nu}-\frac{T}{2}g_{\mu\nu})u^\mu
u^\nu\geq0,\\\label{15*} &&R_{\mu\nu}k^\mu
k^\nu=(T_{\mu\nu}-\frac{T}{2}g_{\mu\nu})k^\mu k^\nu\geq0.
\end{eqnarray}
In the case of perfect fluid matter, the SEC and NEC given by
(\ref{14*}) and (\ref{15*}) impose the following constraints
$\rho+3p\geq0$ and $\rho+p\geq0$ to be satisfied, while the WEC and
DEC require these bounds $\rho\geq0$ and $\rho\pm p\geq0$,
respectively for consistency purposes.

The concept of energy conditions can be extended to the case of
modified theories of gravity using Raychaudhari equation, a purely
geometrical relation. Thus, its interesting particulars like
focussing of geodesic congruences along with the attractive nature
of gravity can be used to formulate these bounds in any modified
gravitational framework. In such cases, we take the total matter
contents of the universe behaving as perfect fluid and consequently,
the respective conditions can be defined by simply replacing the
energy density and pressure, respectively by an effective energy
density and effective pressure as follows
\begin{eqnarray}\nonumber
\textbf{NEC}:\quad&&\rho^{eff}+p^{eff}\geq0,\\\nonumber
\textbf{SEC}:\quad&&\rho^{eff}+p^{eff}\geq0,\quad\rho^{eff}+3p^{eff}\geq0,\\\nonumber
\textbf{WEC}:\quad&&\rho^{eff}\geq0,\quad\rho^{eff}+p^{eff}\geq0,\\\label{15**}
\textbf{DEC}:\quad&&\rho^{eff}\geq0,\quad\rho^{eff}\pm p^{eff}\geq0.
\end{eqnarray}
In is also interesting to mention here that the violation of these
energy bounds ensure the existence of the ghost instabilities (some
interesting feature of modified gravity that support the cosmic
acceleration due to DE).

\subsection{Energy Conditions in modified $f(T)$ Gravity}

Using Eqs.(\ref{9}) and (\ref{10}), the energy conditions for $f(T)$
gravity with non-minimal torsion-matter coupling are obtained as
\begin{eqnarray}\nonumber
\textbf{NEC}&:&\\\label{11}
{\rho}_{eff}+p_{eff}&=&\frac{(\rho+p)[1+\lambda(f_2+12H^2f'_2)]}{1+f'_1-12H^2f''_1
-16\pi{G}\lambda\rho(f'_2-12H^2f''_2)}\geqslant0,\\\nonumber
\textbf{WEC}&:&\\\label{12}
{\rho}_{eff}&=&[1+\lambda(f_2+12H^2f'_2)]\rho-\frac{1}{16\pi{G}}(f_1+12H^2f'_1)\geqslant0
, \\\nonumber &&{\rho}_{eff}+p_{eff}\geqslant0,\\\nonumber
\textbf{SEC}&:&\\\nonumber
{\rho}_{eff}+3p_{eff}&=&-2[1+\lambda(f_2+12H^2f'_2)]\rho+\frac{1}{8\pi{G}}(f_1+12H^2f'_1)
\\\label{13}&+&\frac{3(\rho+p)[1+\lambda(f_2+12H^2f'_2)]}{1+f'_1-12H^2f''_1
-16\pi{G}\lambda\rho(f'_2-12H^2f''_2)}\geqslant0, \\\nonumber
&&{\rho}_{eff}+p_{eff}\geqslant0, \\\nonumber
\textbf{DEC}&:&\\\nonumber
{\rho}_{eff}-p_{eff}&=&2[1+\lambda(f_2+12H^2f'_2)]\rho-\frac{1}{8\pi{G}}(f_1+12H^2f'_1)
\\\label{14}&-&\frac{(\rho+p)[1+\lambda(f_2+12H^2f'_2)]}{1+f'_1-12H^2f''_1
-16\pi{G}\lambda\rho(f'_2-12H^2f''_2)}\geqslant0, \\\nonumber
&&{\rho}_{eff}+p_{eff}\geqslant0,\quad {\rho}_{eff}\geqslant0.
\end{eqnarray}
The inequalities (\ref{11})-(\ref{14}) represent the null, weak,
strong and dominant energy conditions in the context of $f(T)$
theory with nonminimal torsion-matter coupling for FRW spacetime. In
the following, we consider some specific functional forms for the
Lagrangian (\ref{1}) to develop constraints under these conditions.
Harko et al. (2014) presented some viable models of $f(T)$ gravity
with nonminimal torsion-matter coupling and discuss different
evolutionary phases depending on the coupling and free parameters.
It is shown that different model parameters can result in de Sitter,
DE dominated and accelerating phases.

\subsection{$f_1(T)=-\Lambda+\alpha_1T^2$, $f_2(T)=\beta_1T^2$}

In the first place, we consider the model (Harko et al. 2014)
\begin{eqnarray}\label{14a}
f_1(T)=-\Lambda+\alpha_1T^2,\quad f_2(T)=\beta_1T^2,
\end{eqnarray}
where $\alpha_1$, $\beta_1$ are model parameters and $\Lambda>0$ is
a constant. These functions involve quadratic contribution from $T$
and appear as corrections to teleparallel theory. The derivatives of
these functions are defined as $f'_1=2\alpha_1T$, $f''_1=2\alpha_1$,
$f'_2=2\beta_1T$ and $f''_2=2\beta_1$. Since torsion is defined in
terms of Hubble parameter so we can change the functional dependence
from $T$ to $H$ as $f_1(T){\equiv}f_1(H)=-\Lambda+\alpha{H}^4$ and
$f_2(T){\equiv}f_2(H)=\beta{H}^4$, where $\alpha=36\alpha_1$,
$\beta=36\beta_1$. For the derivatives of $f_1$ and $f_2$, we have
$f'_1(H)=-\alpha{H}^2/3$, $f''_1(H)=\alpha/18$,
$f'_2(H)=-\beta{H}^2/3$ and $f''_2(H)=\beta/18$.

In FRW background, the constraints to fulfil the WEC energy
condition ($\rho_{eff}\geqslant0$, $\rho_{eff}+p_{eff}\geqslant0$)
for such model can be represented as
\begin{eqnarray}\label{15}
\rho(1-3\lambda\beta{H}^4)+\frac{1}{2}(\Lambda+3\alpha{H}^4)\geqslant0,
\\\label{16}
(\rho+p)\{(1-3\lambda\beta{H}^4)/(1+(2\lambda\beta\rho-\alpha)H^2)\}\geqslant0.
\end{eqnarray}
In terms of present day value of $H$, the inequality (\ref{15}) can
be satisfied if we set $\beta<0$ with $(\Lambda, \alpha,
\lambda)>0$. For the inequality (\ref{16}), if one set $\beta<0$
then it requires $(2\lambda\beta\rho+\alpha)<1$. It can also be
satisfied by choosing the parameters such that
$2\lambda\beta\rho>\alpha$ and $1>3\lambda\beta{H}^4$.

To be more explicit about the validity of these inequalities, we
consider the power law cosmology
\begin{equation}\label{17}
a(t)=a_0t^m,
\end{equation}
where $m$ is a positive real number. If $0<m<1$, then the required
power law solution is decelerating, while for $m>1$, it exhibits
accelerating behavior. We set $m>1$ with $\rho=\rho_0t^{-3m}$. To
explore the validity of inequalities (\ref{15}) and (\ref{16}), we
present the evolution of WEC for various values of parameters. In
Figure \textbf{1}, we show the variation of inequalities (\ref{15})
and (\ref{16}) versus $t$ and $\lambda$. We fix the model parameters
$\alpha$ and $\beta$ to show the evolution for different values of
$\lambda$. It is shown that WEC can be satisfied if $\lambda>0$ with
$\alpha=0.01$ and $\beta=0.02$. In Figure \textbf{2}, we fix
$\lambda$ and vary the parameter $\beta$, the plots show that WEC
can be met for both positive and negative values of $\beta$. We also
present the evolution of WEC for various values of parameter
$\alpha$ in Figure \textbf{3}. The left plot shows that inequality
(\ref{15}) is satisfied only for positive values of parameter
$\alpha$ and violates for negative values whereas in right plot it
can be met for all values of $\alpha$. The inequality (\ref{16})
represents the NEC in non-minimally coupled $f(T)$ gravity for the
model (\ref{14a}).
\begin{figure}
\centering \epsfig{file=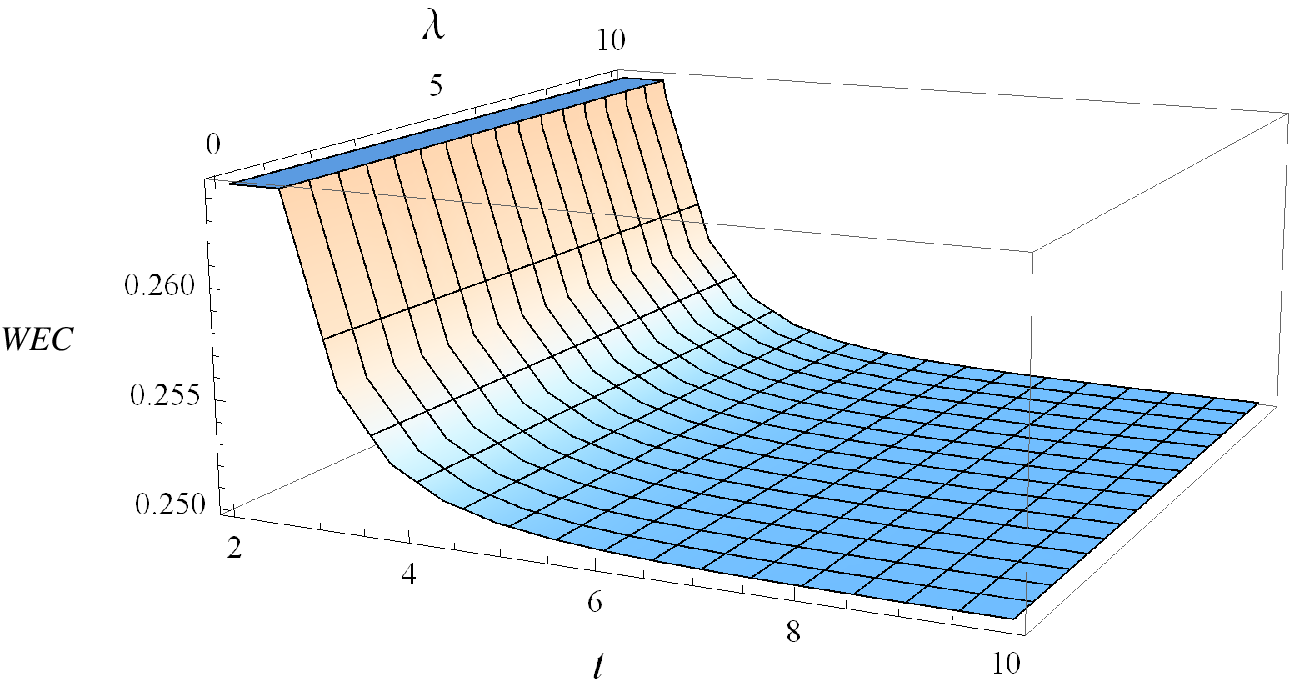, width=.49\linewidth,
height=1.6in}\epsfig{file=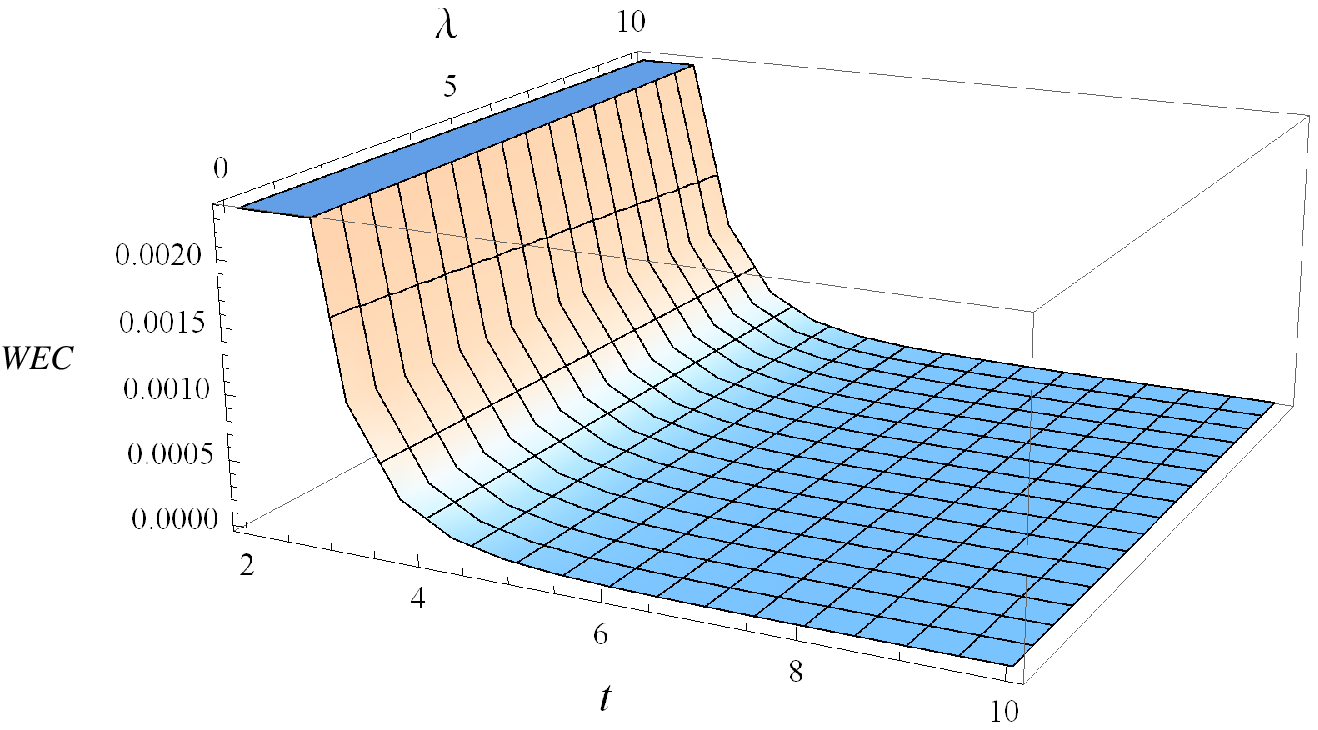, width=.49\linewidth,
height=1.6in}\caption{Evolution of WEC versus $t$ and $\lambda$ for
$\alpha=0.01$, $\beta=0.02$, $\Lambda=.5$ and $m=2$. The left and
right plots correspond to inequalities (\ref{15}) and (\ref{16})
respectively.}
\end{figure}
\begin{figure}
\centering \epsfig{file=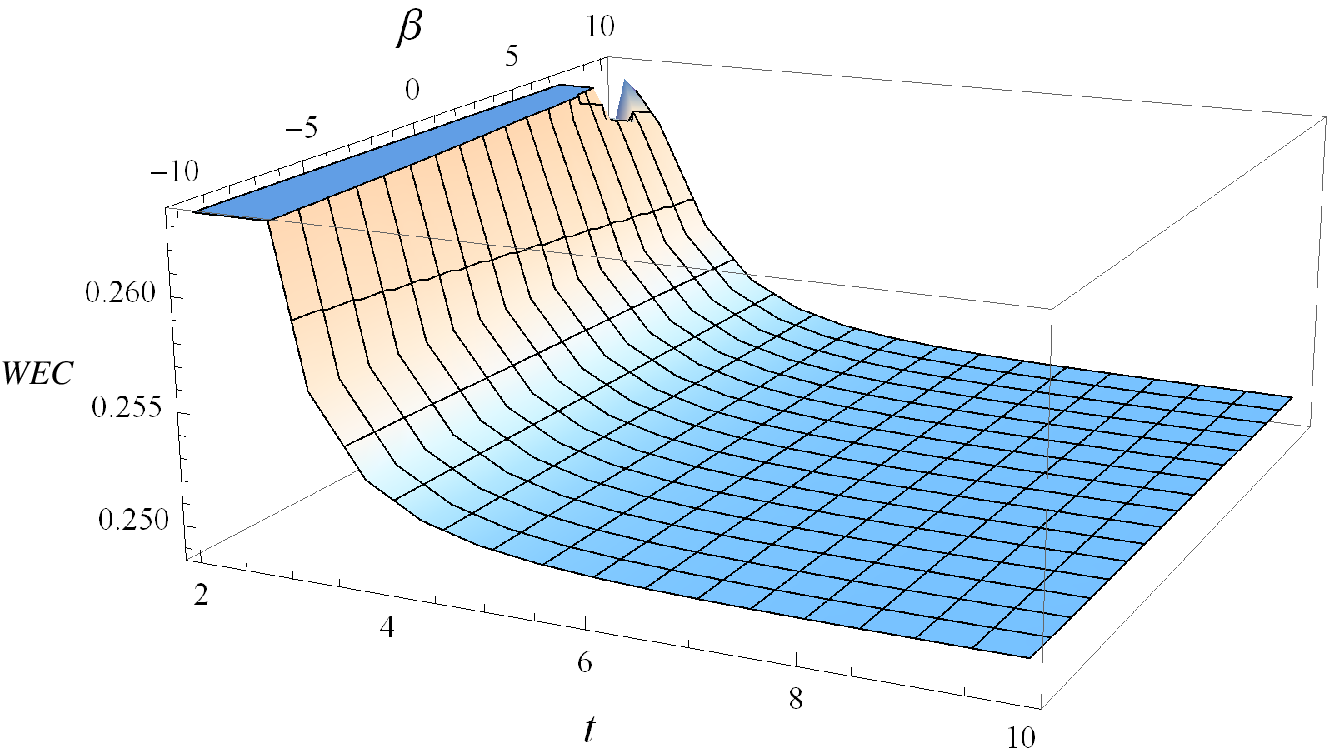, width=.49\linewidth,
height=1.6in}\epsfig{file=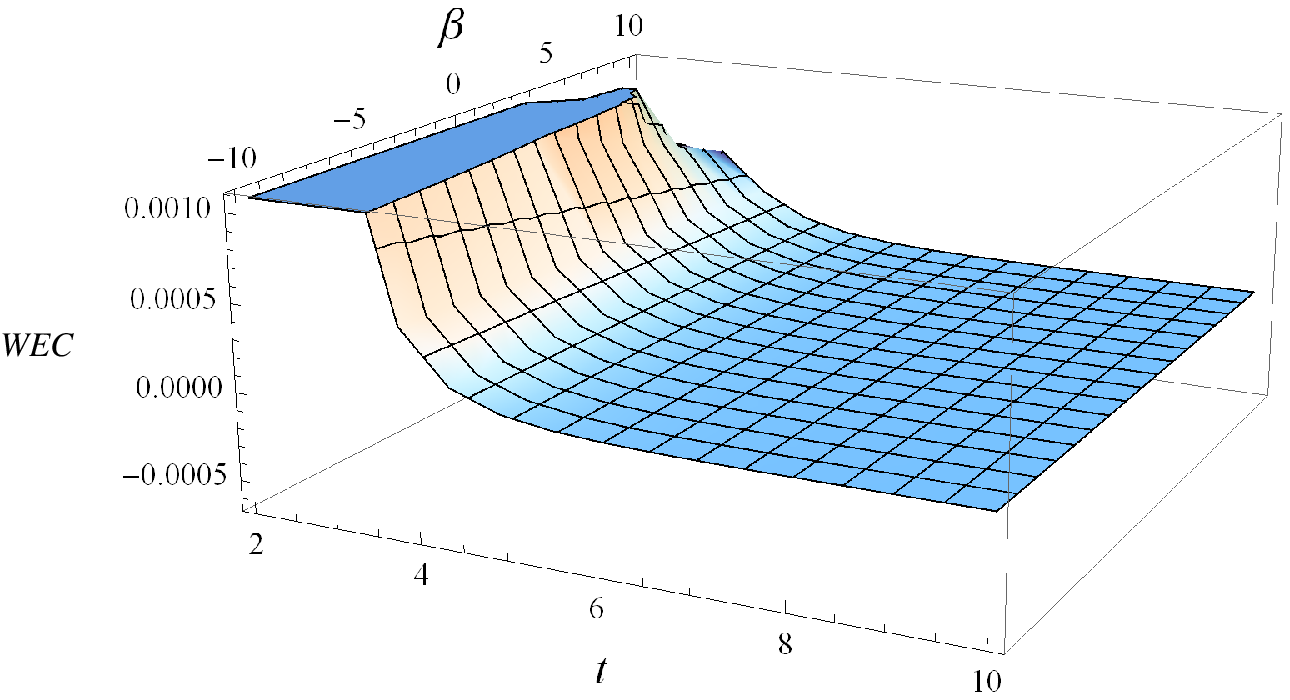, width=.49\linewidth,
height=1.6in}\caption{Evolution of WEC versus $t$ and $\beta$ for
$\alpha=0.01$, $\lambda=.1$, $\Lambda=.5$ and $m=2$. The left and
right plots correspond to inequalities (\ref{15}) and (\ref{16})
respectively.}
\end{figure}
\begin{figure}
\centering \epsfig{file=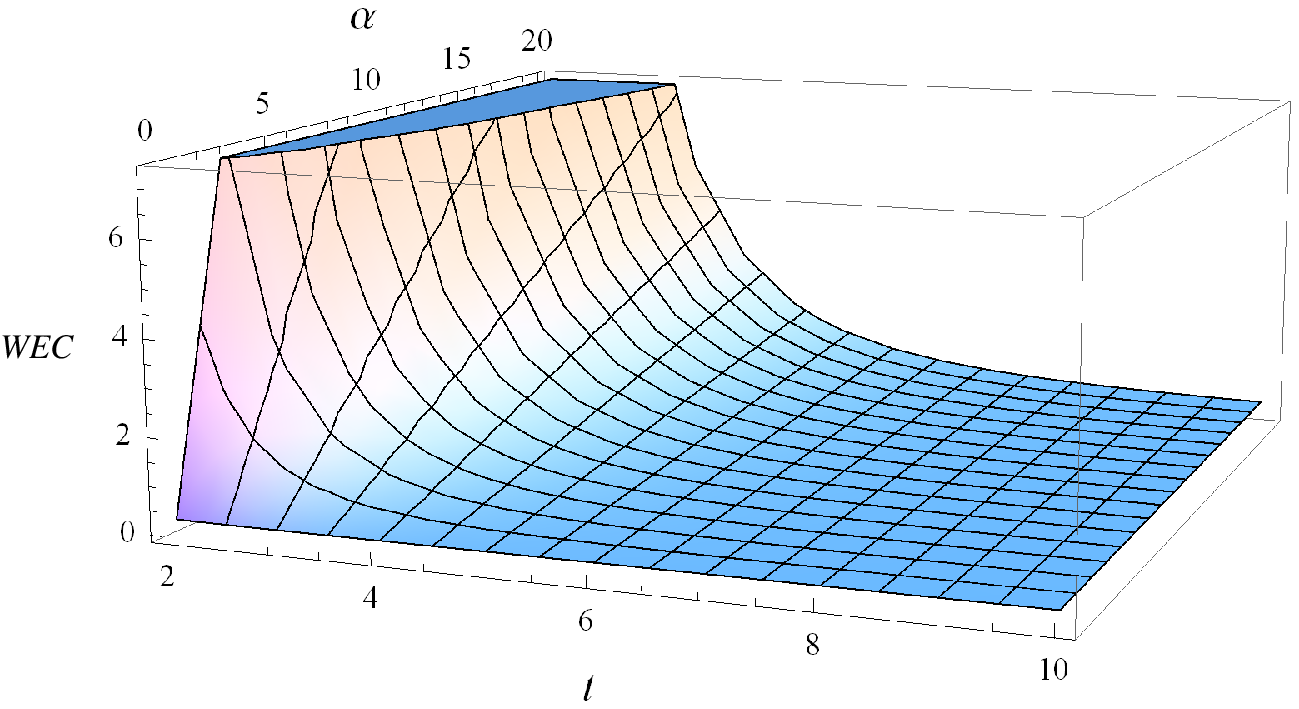, width=.49\linewidth,
height=1.6in}\epsfig{file=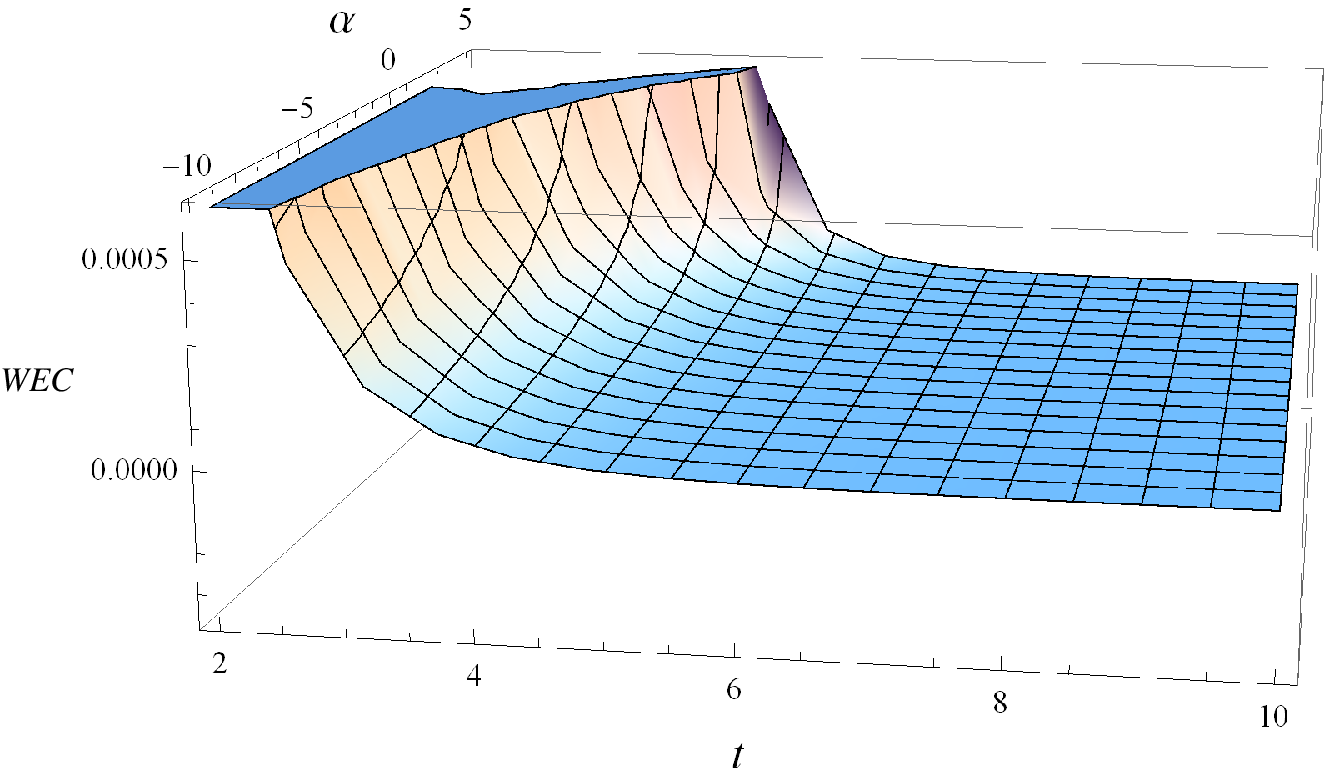, width=.49\linewidth,
height=1.6in}\caption{Evolution of WEC versus $t$ and $\alpha$ for
$\beta=.02$, $\lambda=.1$, $\Lambda=.5$ and $m=2$. The left and
right plots correspond to inequalities (\ref{15}) and (\ref{16})
respectively.}
\end{figure}
The constraints to fulfill the SEC and DEC in power law cosmology
for above model can be found from inequalities (\ref{13}) and
(\ref{14}). We also show the evolution of SEC and DEC in Figures
\textbf{4} and \textbf{5} for the $f(T)$ model (\ref{14a}). In plot
\textbf{4(a)}, we show the variation of SEC versus $\lambda$ and
$t$. Here, SEC can be met for $\lambda>0$ if $\alpha<0$. The
evolution of SEC is also presented for various values of $\alpha$
and $\beta$. We find that SEC holds for $\alpha<0$ with fixed values
of $\lambda$, $\beta$ and $\Lambda$ as shown in Figure
\textbf{4(b)}. In Figure \textbf{4(c)}, we vary $\beta$ for fixed
values of other parameters and find that SEC can be satisfied for
any value of $\beta$. We also present the evolution of DEC for
different values of parameters in Figure \textbf{5}. In this case
DEC can be satisfied for $\lambda>0$ with $\alpha>0$. The evolution
for the parameter $\alpha$ and $\beta$ is also shown in Figures
\textbf{5(b)} and \textbf{5(c)}.
\begin{figure}
\centering \epsfig{file=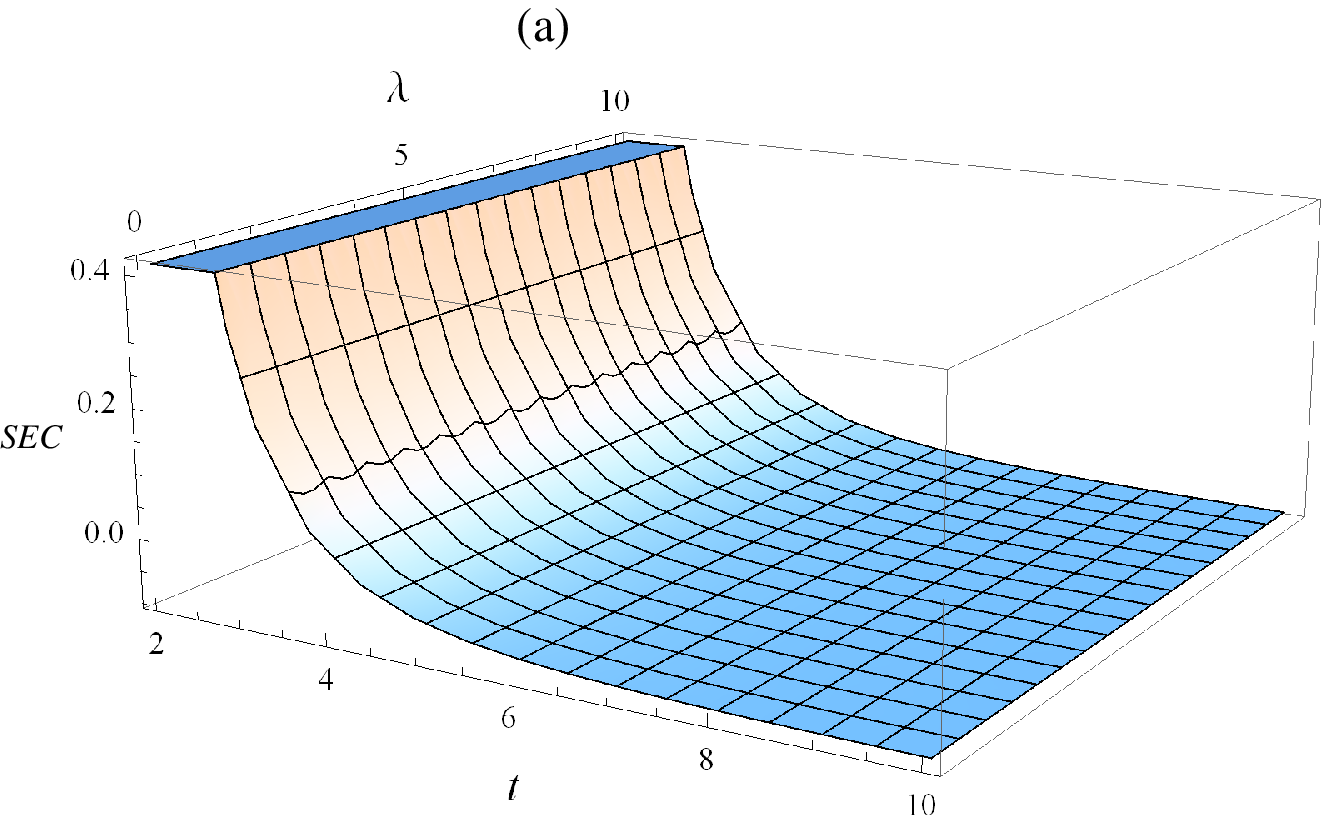, width=.49\linewidth,
height=1.8in}\epsfig{file=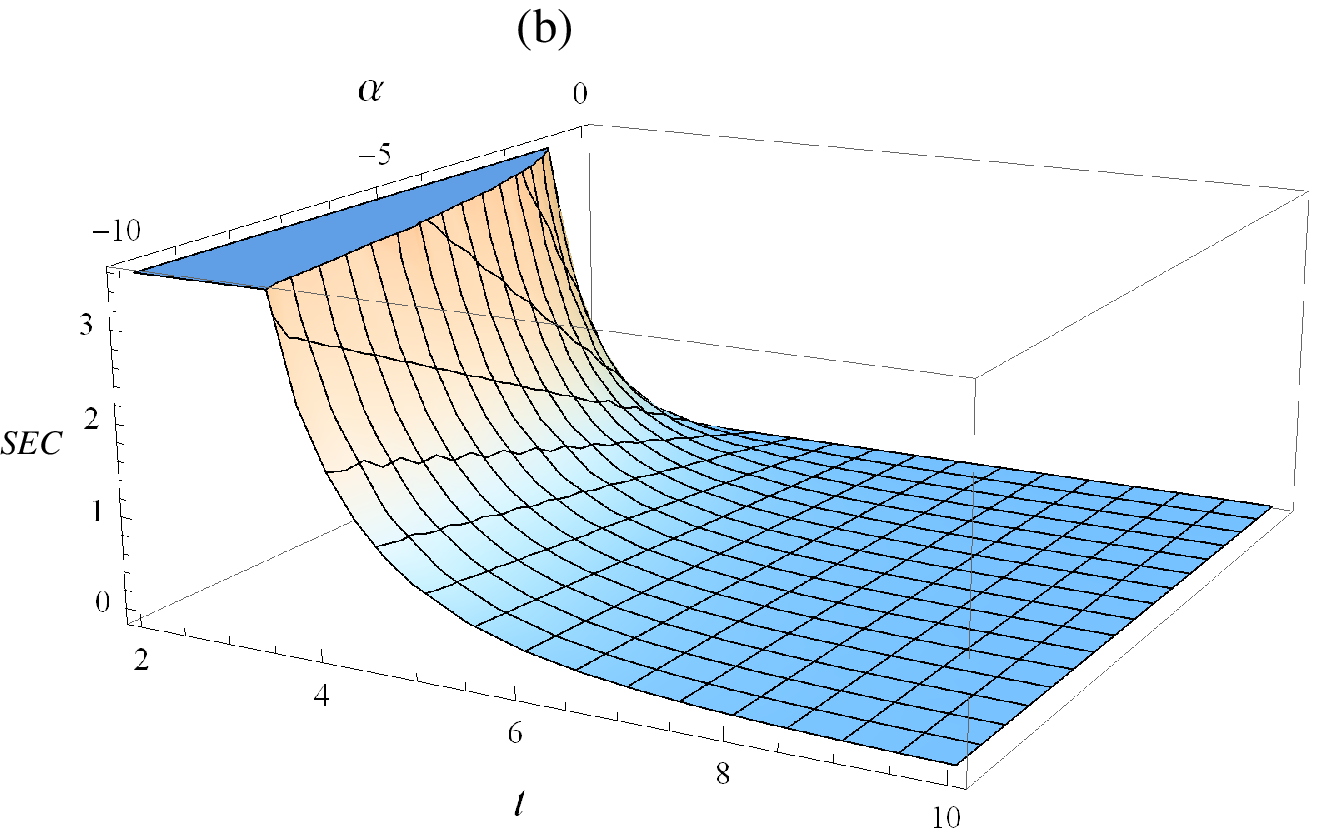, width=.49\linewidth,
height=1.8in}\\\epsfig{file=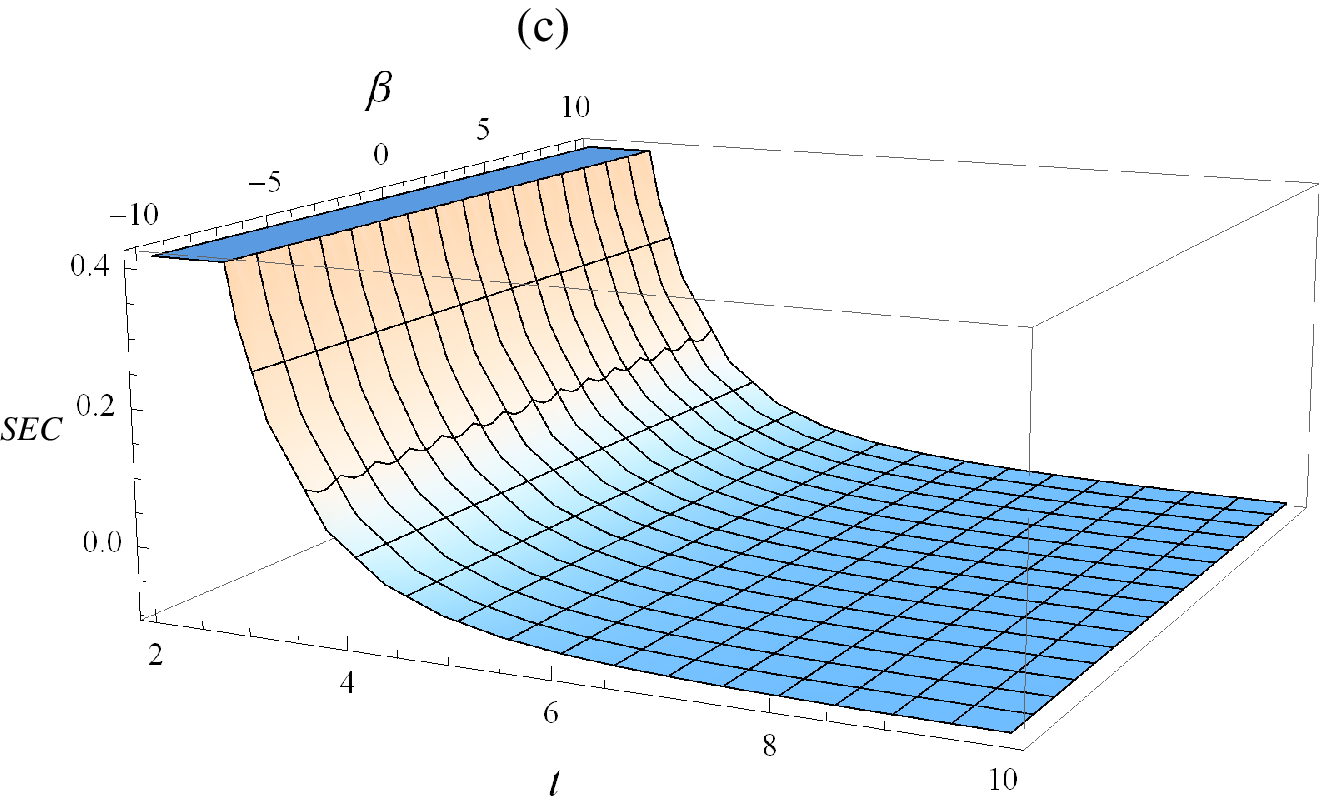, width=.49\linewidth,
height=1.8in}\caption{Evolution of SEC (a) versus $t$ and $\lambda$
for $\alpha=-0.06$, $\beta=.02$, $\Lambda=.1$ and $m=2$ (b) versus
$t$ and $\alpha$ for $\beta=0.02$ (c) versus $t$ and $\beta$ for
$\alpha=-0.06$.}
\end{figure}
\begin{figure}
\centering \epsfig{file=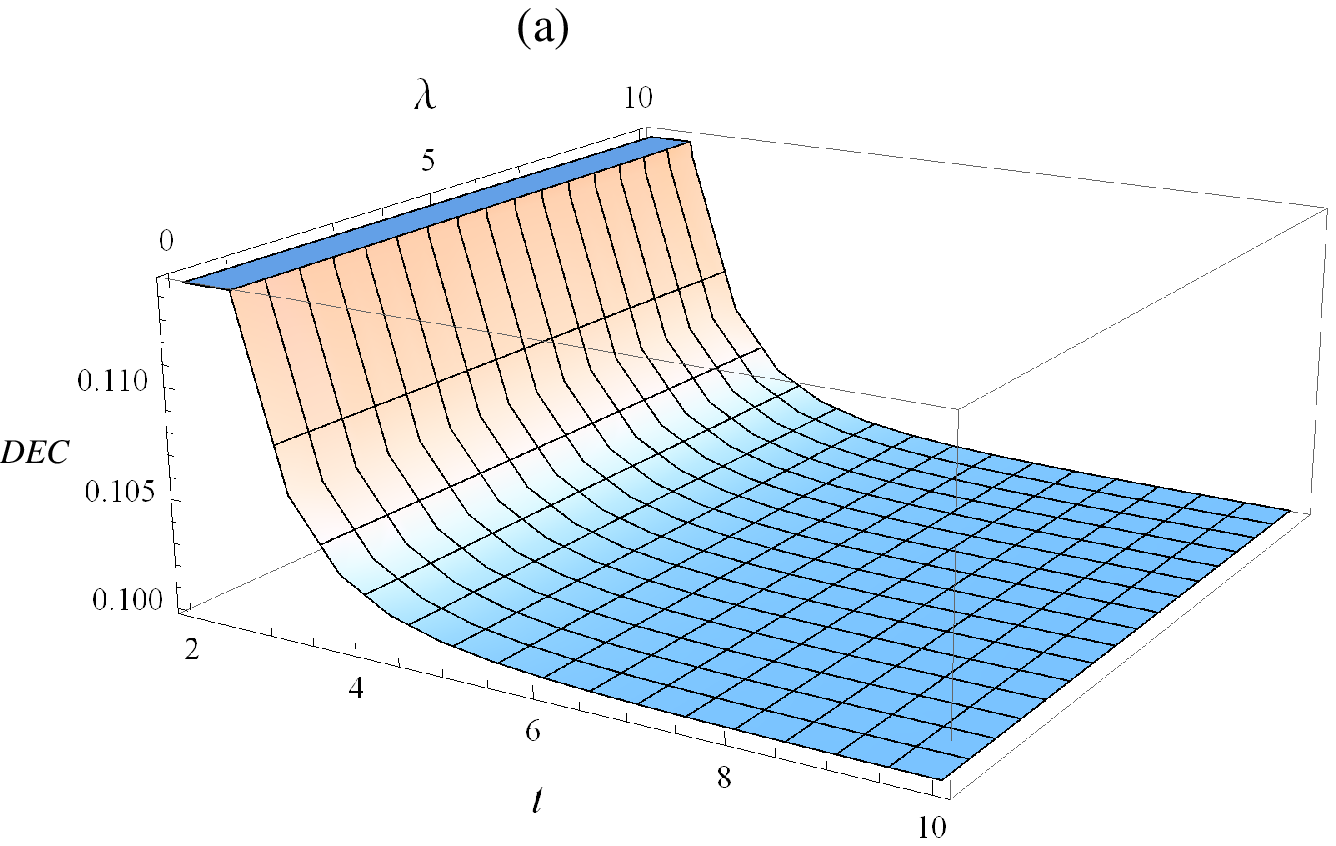, width=.49\linewidth,
height=1.8in}\epsfig{file=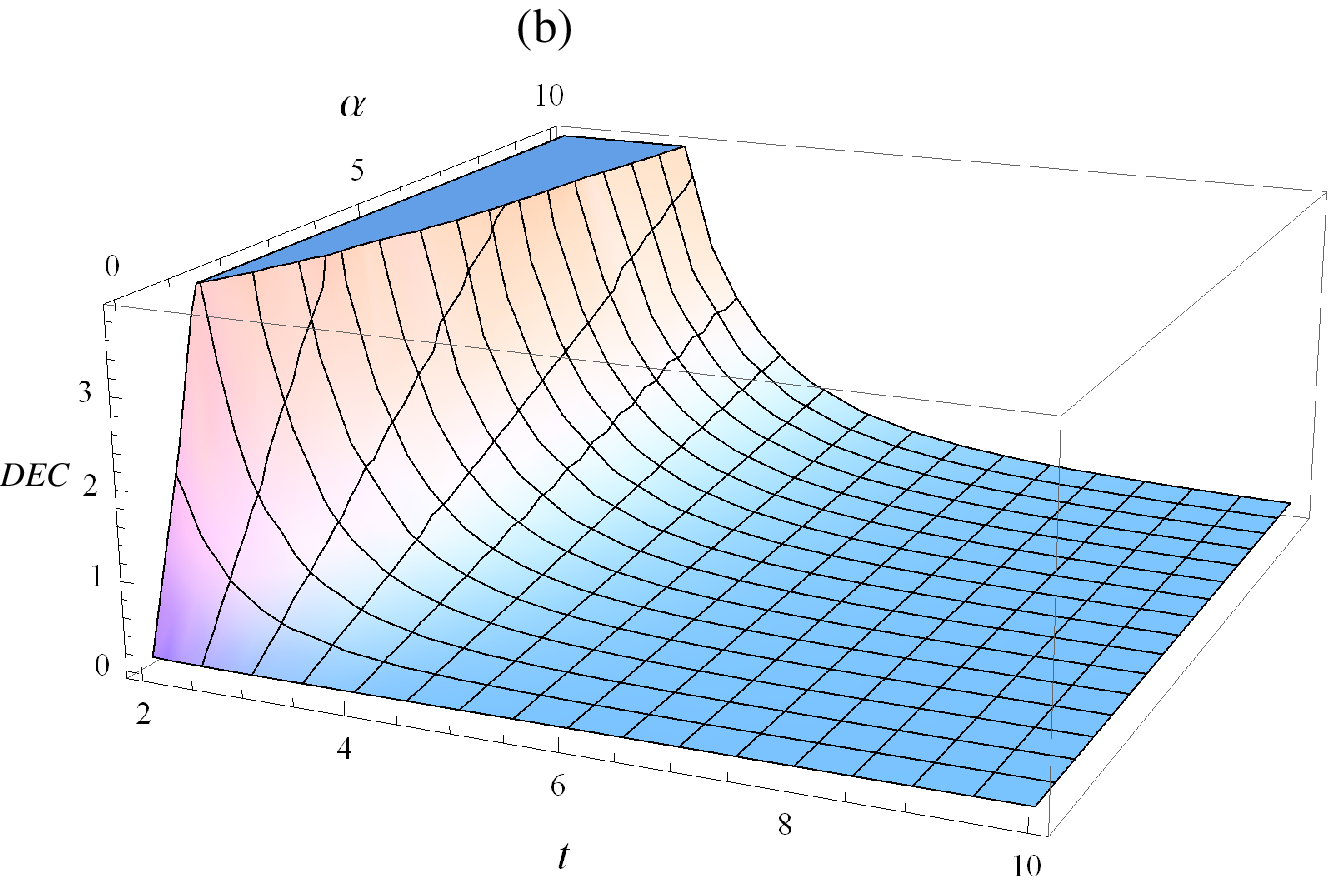, width=.49\linewidth,
height=1.8in}\\\epsfig{file=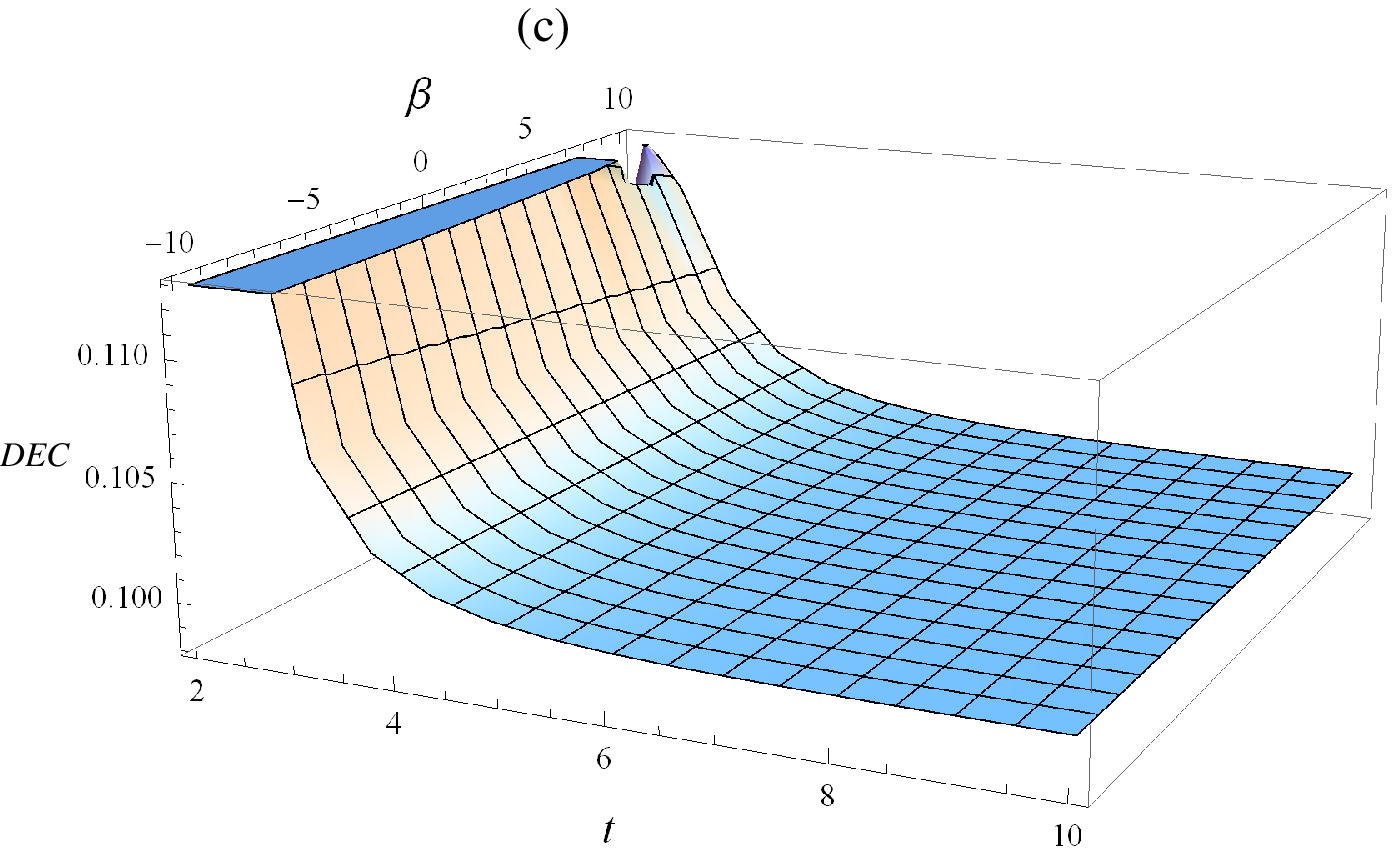, width=.49\linewidth,
height=1.8in}\caption{Evolution of DEC (a) versus $t$ and $\lambda$
for $\alpha=0.01$, $\beta=0.02$, $\Lambda=.1$ and $m=2$ (b) versus
$t$ and $\alpha$ for $\beta=0.02$ and $\lambda=0.1$ (c) versus $t$
and $\beta$ for $\alpha=0.01$.}
\end{figure}

\subsection{$f_1(T)=-\Lambda$, $f_2(T)=\alpha_2T+\beta_2T^2$}

In second example, we consider the model defined by the following
functions (Harko et al. 2014)
\begin{eqnarray}\label{18}
f_1(T)=-\Lambda,\quad f_2(T)=\alpha_2T+\beta_2T^2,
\end{eqnarray}
where $\alpha_2$ and $\beta_2$ are parameters for the model
(\ref{18}). We express the functions $f_1$ and $f_2$ in terms of $H$
as $f_1(H)=-\Lambda$, $f_2(H)=\gamma{T}+\delta{T}^2$, where
$\gamma=-6\alpha_2$ and $\delta=36\beta_2$. Similarly, the
derivatives of $f_1$ and $f_2$ are given by $f'_1(H)=f''_1(H)=0$,
$f'_2(H)=-\gamma/6-\delta{H}^2/3$ and $f''_2(H)=\delta/18$.

The WEC ($\rho_{eff}\geqslant0$, $\rho_{eff}+p_{eff}\geqslant0$) for
the model (\ref{18}) requires the following inequalities to be
satisfied
\begin{eqnarray}\label{19}
\rho+\Lambda/2-\lambda\rho(\gamma{H}^2+3\delta{H}^4)\geqslant0,
\\\label{20}
(\rho+p)\{[1-\lambda(\gamma{H}^2+3\delta{H}^4)]/[1+2\lambda\rho
(\gamma/6+\delta{H}^2)]\}\geqslant0.
\end{eqnarray}
It can be seen that above inequalities require $(\gamma, \delta)<0$
to satisfy the WEC in this case. We present the evolution of
constraints (\ref{19}) and (\ref{20}) for different choices of
parameters in Figures \textbf{6}-\textbf{8}. In Figure \textbf{6},
we develop the range for coupling parameter $\lambda$ to satisfy the
WEC. WEC can be satisfied for $\lambda>0$ if $\gamma=-0.1$,
$\delta=-0.2$, $\Lambda=1$ and $m=2$.
\begin{figure}
\centering \epsfig{file=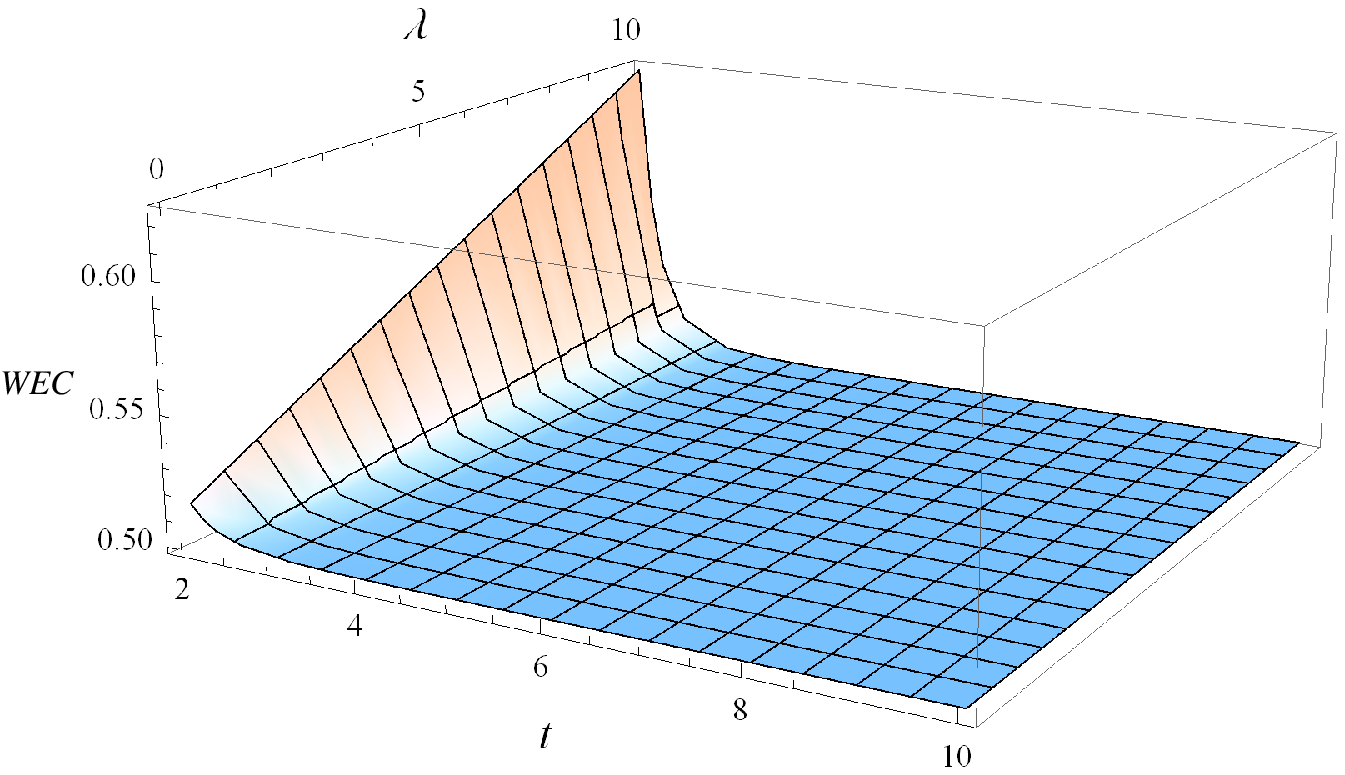, width=.49\linewidth,
height=1.8in}\epsfig{file=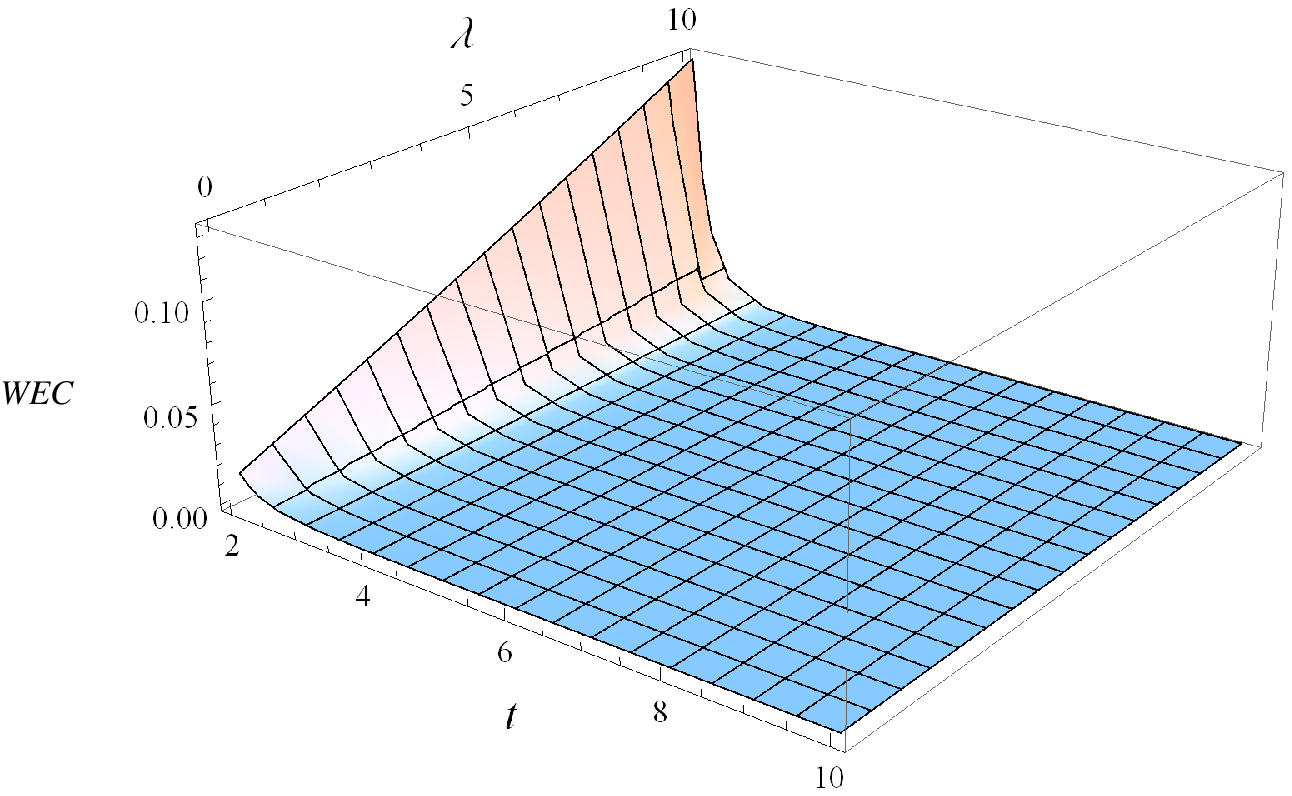, width=.49\linewidth,
height=1.8in}\caption{Evolution of WEC versus $t$ and $\lambda$ for
$\gamma=-0.1$, $\delta=-0.2$ and $m=2$ for the model (\ref{18}). The
left and right plots correspond to constraints (\ref{19}) with
$\Lambda=1$ and (\ref{20}) respectively.}
\end{figure}
In Figure \textbf{7}, evolution of WEC is shown for different values
of parameter $\gamma$. WEC can be met for $\gamma=\{-10,...,10\}$ if
we set $\delta=-0.2$, $\lambda=0.01$ and $m=2$. We also present the
validity of WEC for $\delta=\{-10,...,10\}$ with $\gamma=-0.1$,
$\lambda=0.01$ and $m=2$ as shown in Figure \textbf{8}.
\begin{figure}
\centering \epsfig{file=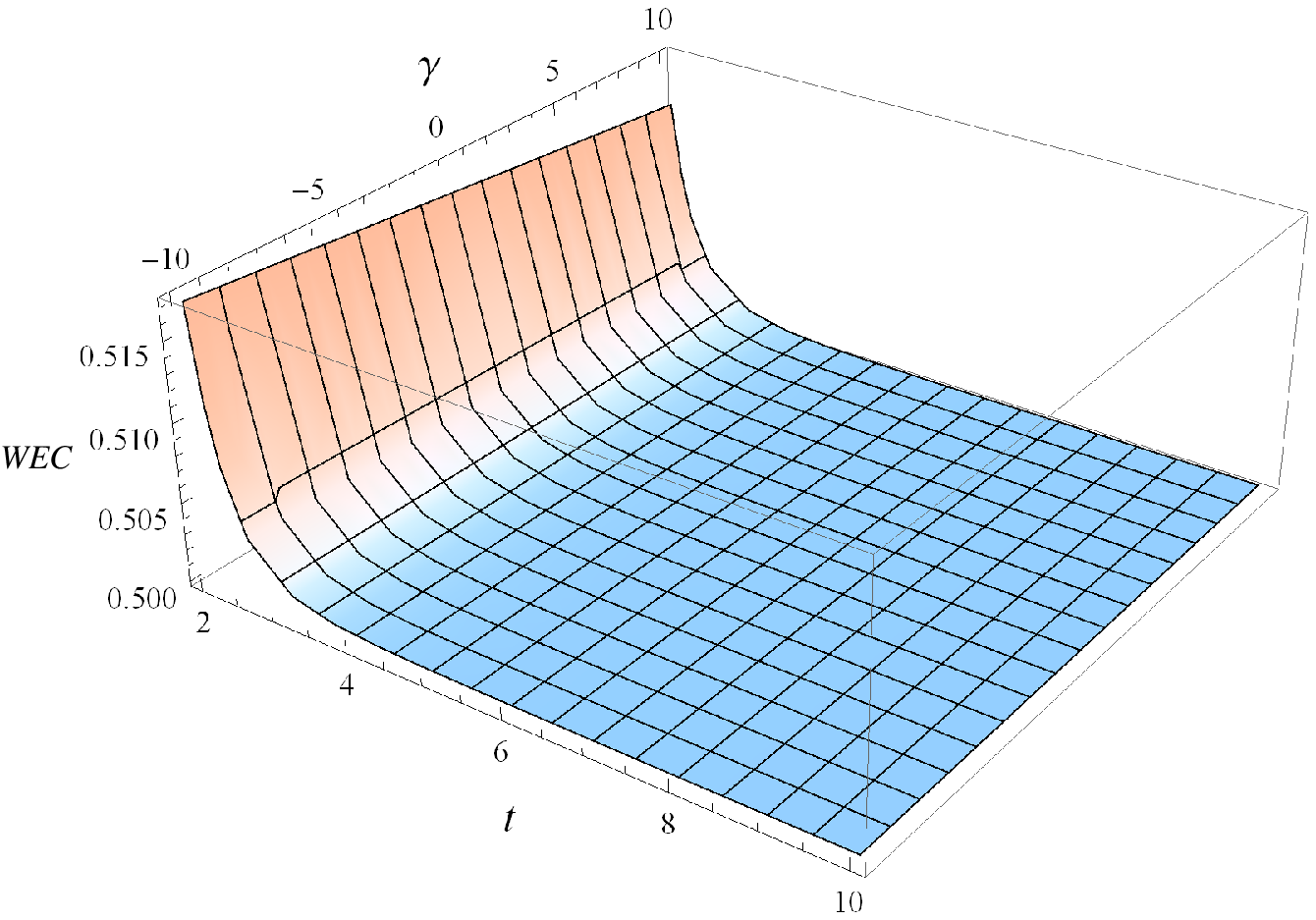, width=.49\linewidth,
height=1.8in}\epsfig{file=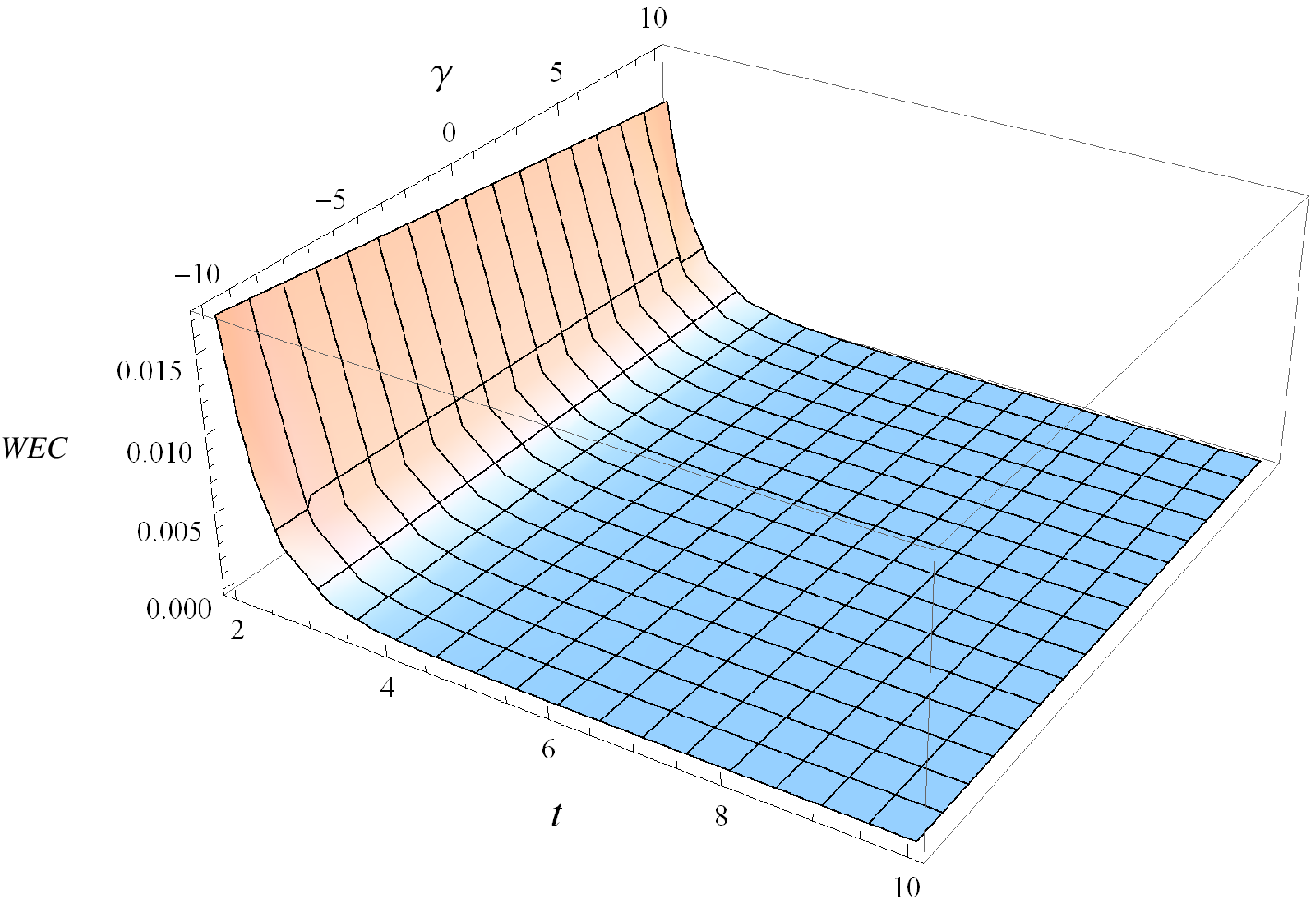, width=.49\linewidth,
height=1.8in}\caption{Evolution of WEC versus $t$ and $\gamma$ for
$\delta=-0.2$, $\lambda=0.01$ and $m=2$ for the model (\ref{18}).
The left and right plots correspond to constraints (\ref{19}) with
$\Lambda=1$ and (\ref{20}) respectively.}
\end{figure}
\begin{figure}
\centering \epsfig{file=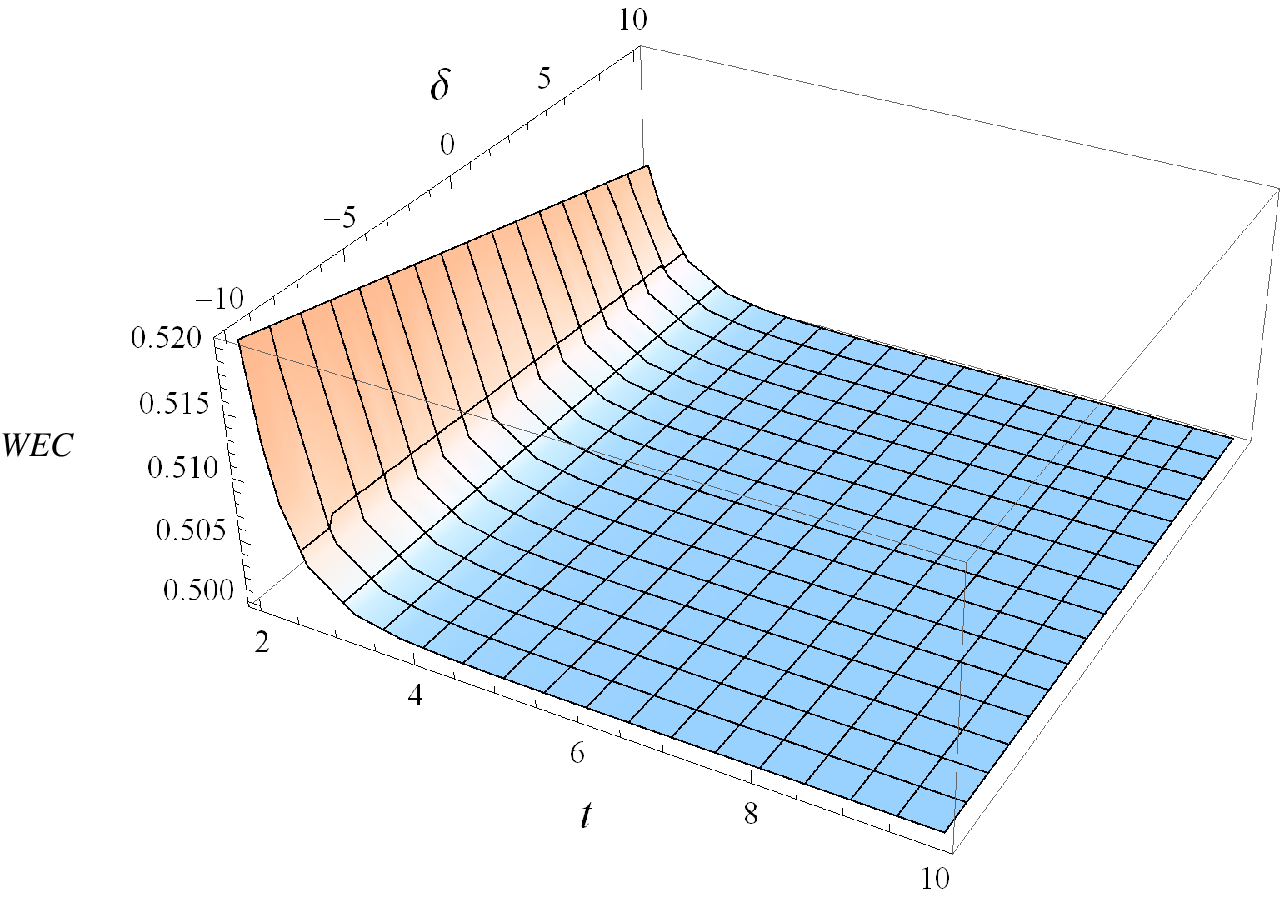, width=.49\linewidth,
height=2.1in}\epsfig{file=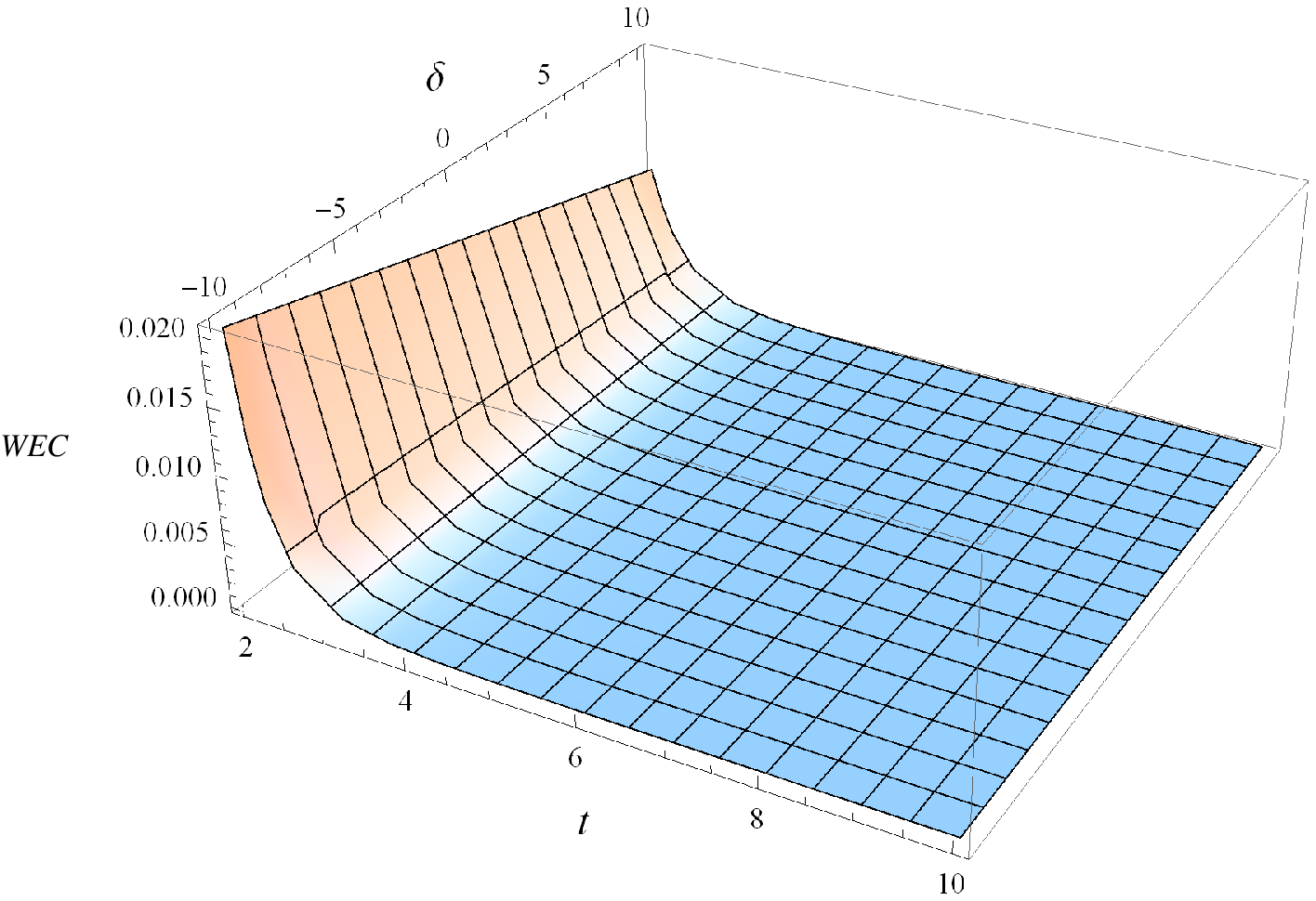, width=.49\linewidth,
height=2.1in}\caption{Evolution of WEC versus $t$ and $\delta$ for
$\gamma=-0.1$, $\lambda=0.01$ and $m=2$ for the model (\ref{18}).
The left and right plots correspond to constraints (\ref{19}) with
$\Lambda=1$ and (\ref{20}) respectively.}
\end{figure}
We also explore different possibilities to validate the SEC and DEC
for the model (\ref{18}). Figure \textbf{9} shows the evolution of
SEC representing the variation for the parameters $\lambda$,
$\gamma$ and $\delta$. The SEC can be satisfied for $\lambda>0$ if
$\gamma=1$ and $\delta=2$ as shown in Figure \textbf{9(a)}. We find
that SEC can hold for $\gamma>0$ but it needs some specific value
for the parameter $\delta$, like $\gamma=\{0,...,10\}$ requires
$\delta=160$ as shown in Figure \textbf{9(b)}. Similarly, one can
fix $\gamma$ to constrain $\delta$ for the validity of SEC. Figure
\textbf{9(c)} depicts the validity of SEC for $\delta=\{0,...,10\}$
with $\gamma=20$.
\begin{figure}
\centering \epsfig{file=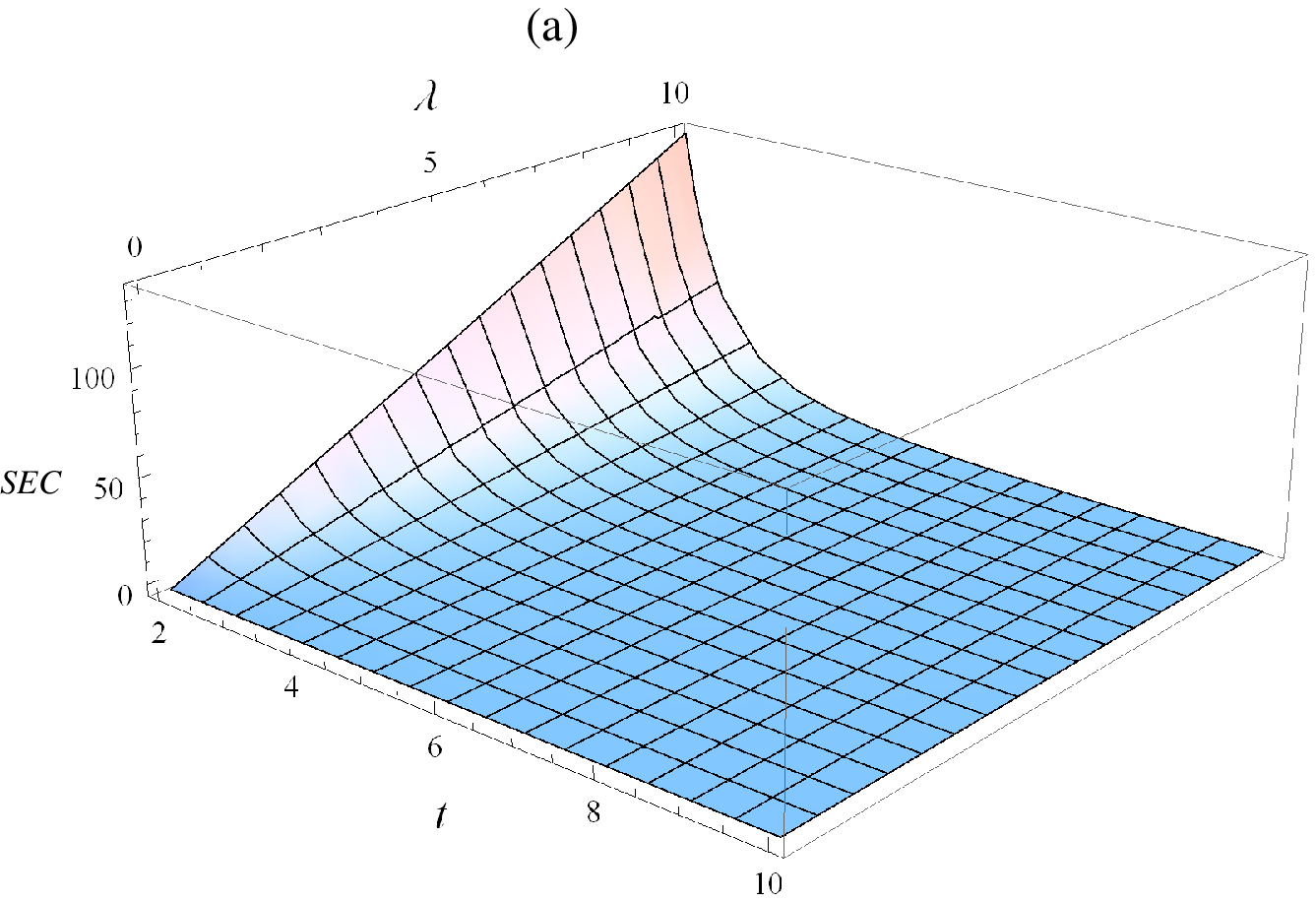, width=.49\linewidth,
height=2.1in}\epsfig{file=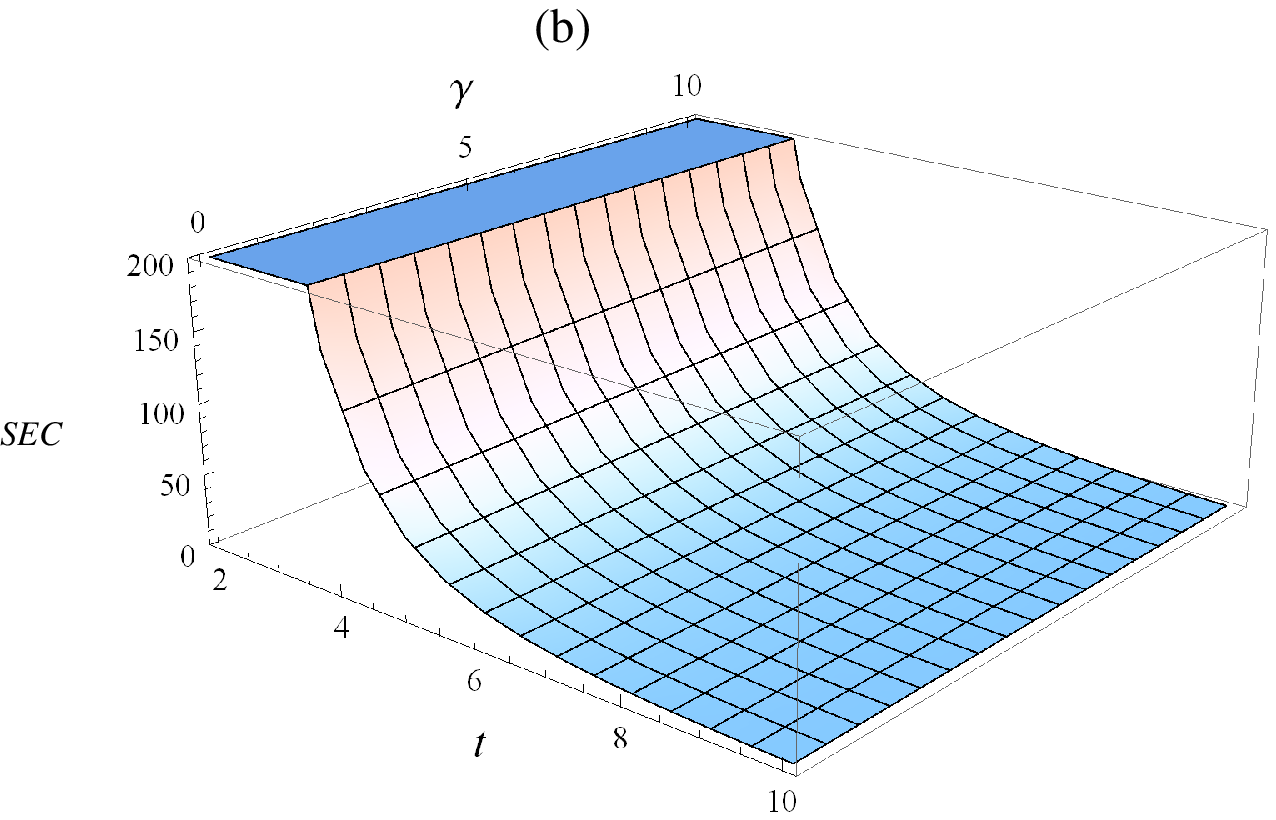, width=.49\linewidth,
height=2.1in}\\\epsfig{file=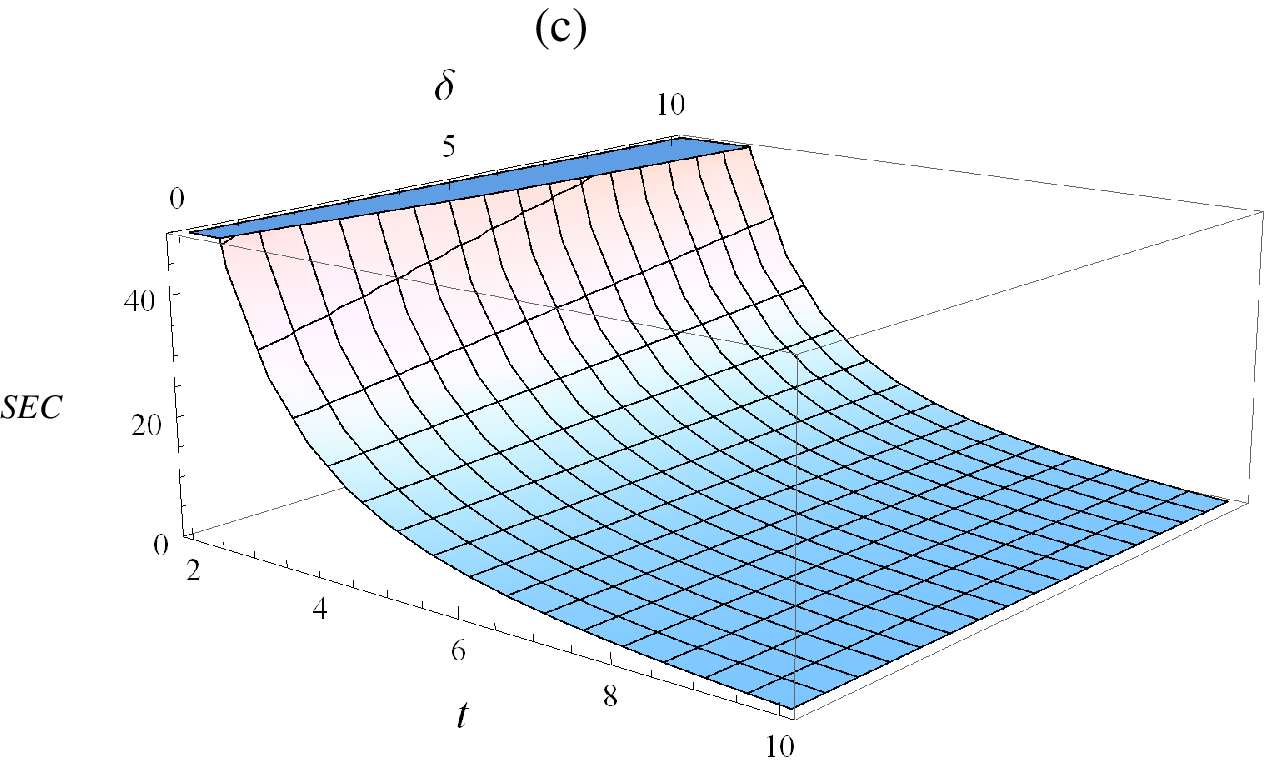, width=.49\linewidth,
height=2.1in}\caption{Evolution of SEC (a) versus $t$ and $\lambda$
for $\gamma=1$, $\delta=2$, $\Lambda=1$ and $m=2$ (b) versus $t$ and
$\gamma$ for $\delta=160$, $\lambda=2$ (c) versus $t$ and $\delta$
for $\gamma=20$.}
\end{figure}
We also explore the validity of DEC in Figure \textbf{10} and
present the respective constraints. One need to set negative values
for $\gamma$ and $\delta$ to validate the DEC for $\lambda>0$ as
shown in plot \textbf{10(a)}. We find that DEC can be met for
negative values of parameters $\gamma$ and $\delta$. We also test
the positive values these parameters as shown in plots
\textbf{10(b)} and \textbf{10(c)}. These plots show that DEC is
satisfied for $\gamma=\{-10,...,10\}$ if $\delta=-12$ and similarly
for $\delta=\{-10,...,10\}$ it requires $\gamma=-24$.
\begin{figure}
\centering \epsfig{file=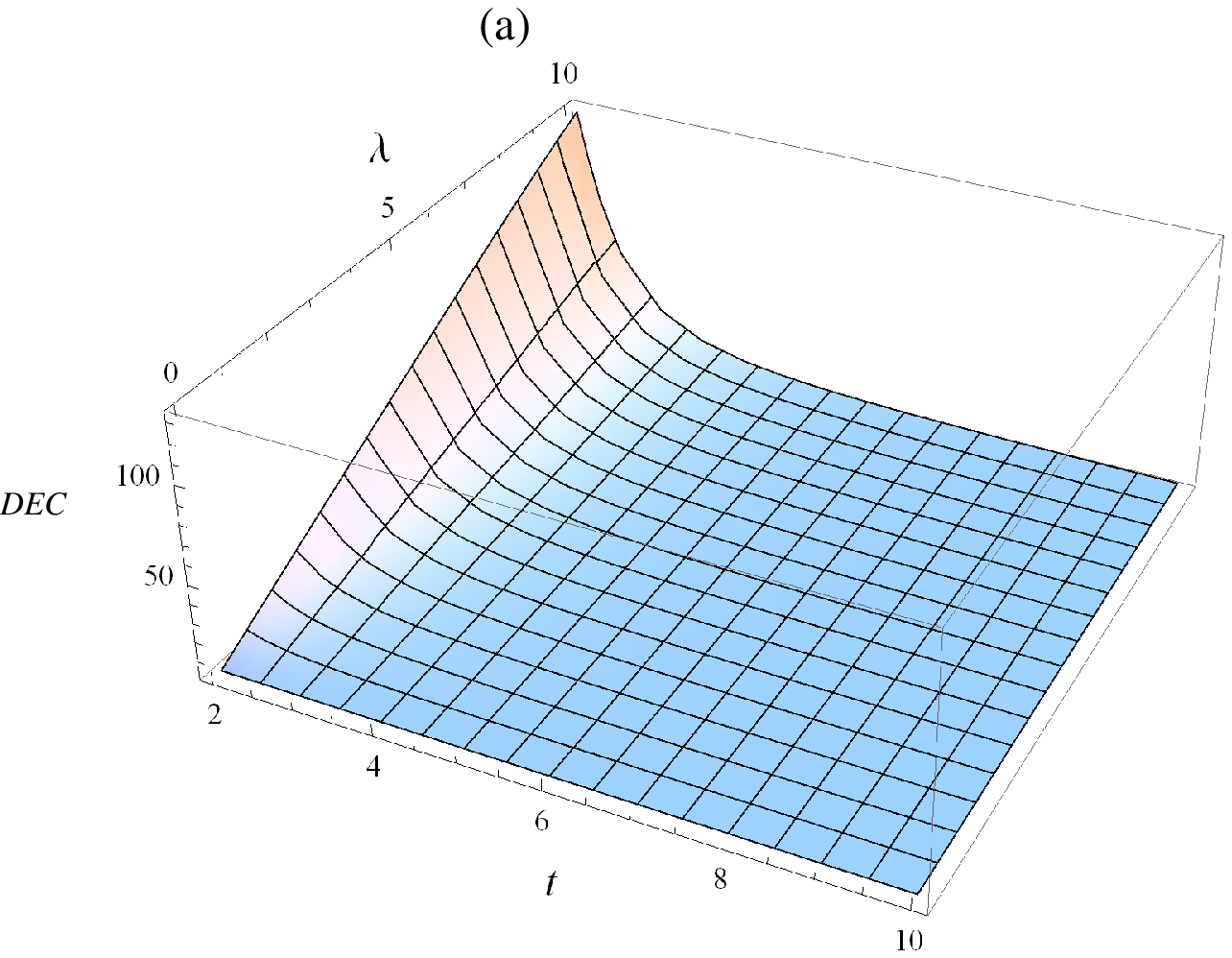, width=.49\linewidth,
height=2.1in}\epsfig{file=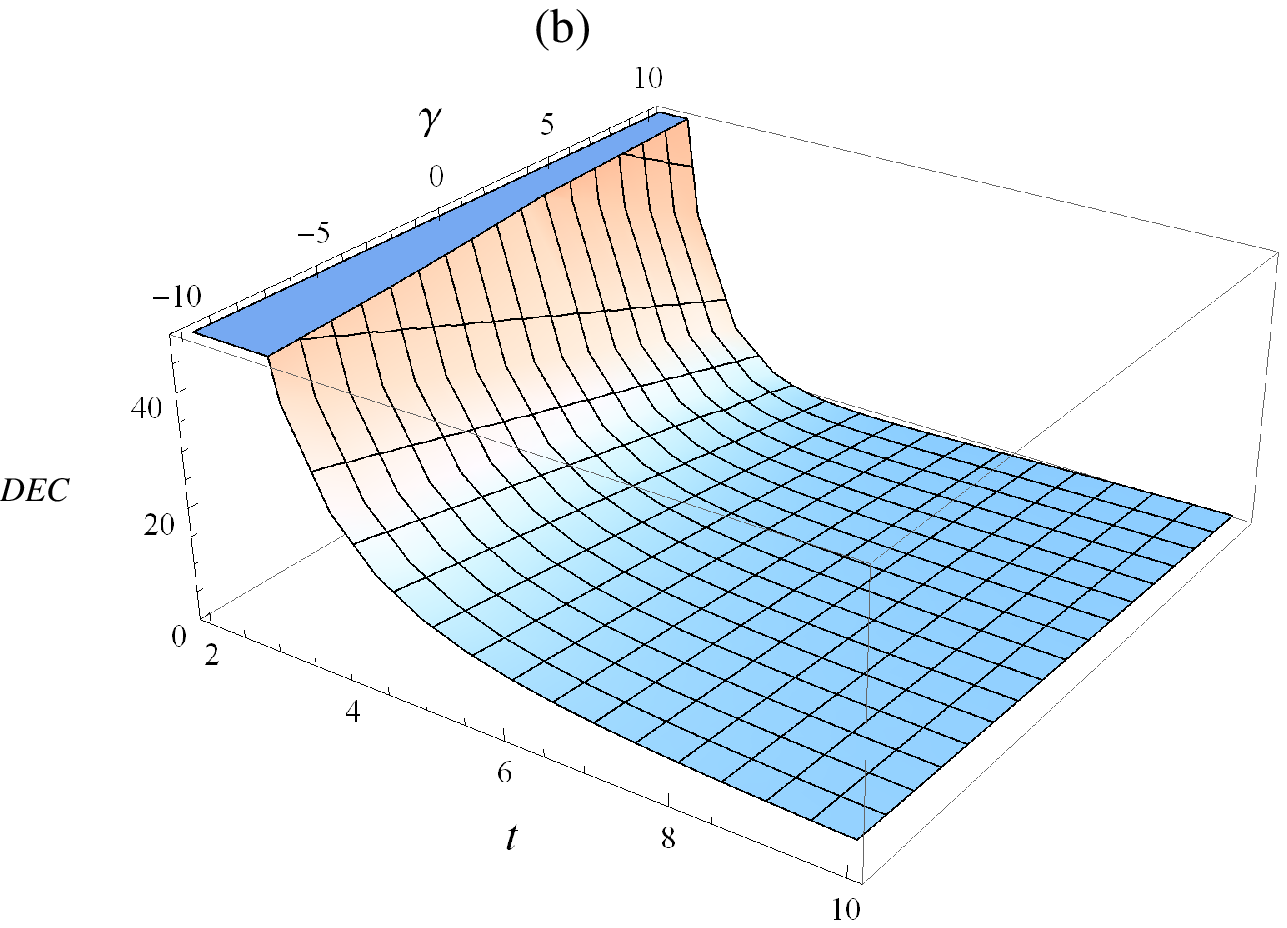, width=.49\linewidth,
height=2.1in}\\\epsfig{file=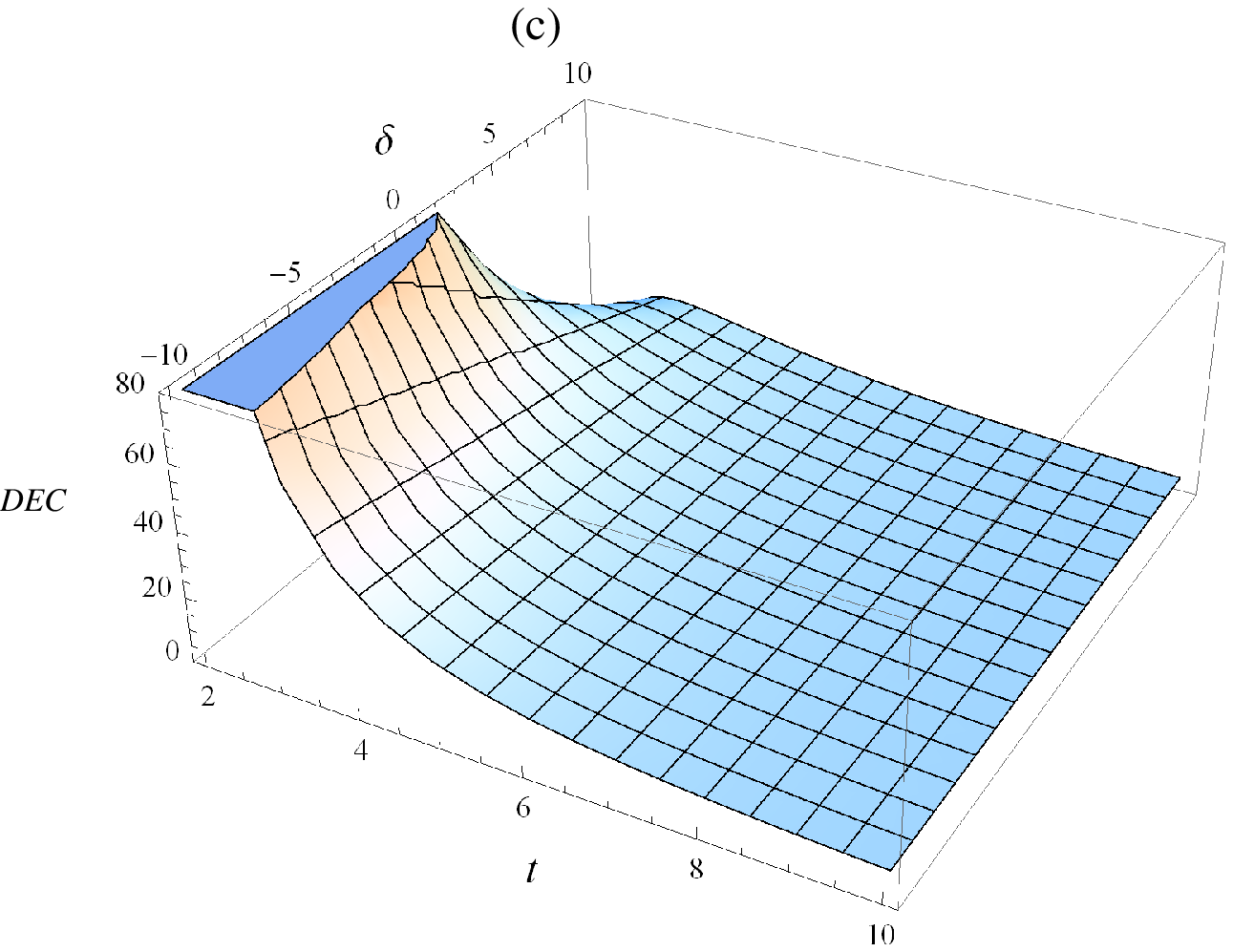, width=.49\linewidth,
height=2.1in}\caption{Evolution of DEC (a) versus $t$ and $\lambda$
for $\gamma=-1$, $\delta=-2$, $\Lambda=1$ and $m=2$ (b) versus $t$
and $\gamma$ for $\delta=-12$, $\lambda=2$ (c) versus $t$ and
$\delta$ for $\gamma=-24$.}
\end{figure}

\section{Conclusions}

Modified theories of gravity have attained significant attention to
explain the observed accelerated cosmic expansion. In this
perspective $f(T)$ gravity has appeared as handy candidate and many
interesting results have been discussed in this theory (Ferraro and
Fiorini 2007; Bamba et al. 2012b; Setare and Darabi 2012; Setare and
Mohammadipour 2012, 2013; Rodrigues et al. 2012, 2013, 2014; Karami
et al. 2013). Recently, an extension of $f(T)$ gravity involving
non-minimal matter torsion coupling is presented in (Harko et al.
2014) which introduces two independent functions of torsion. In such
theories, one can develop different cosmological models depending on
the choice of these functions. The various forms of Lagrangian
raises a question how to constrain such theory on physical grounds.
In this paper, we have developed constraints on specific forms
$f(T)$ by examining the respective energy conditions. We have
derived the energy conditions directly from the effective energy
momentum tensor under the transformation $\rho\rightarrow\rho_{eff}$
and $p_{eff}\rightarrow{p}_{eff}$.

To illustrate how these conditions can constrain the $f(T)$ gravity
with non-minimal matter torsion coupling, we consider two particular
forms of Lagrangian namely, (i) $f_1(T)=-\Lambda+\alpha_1T^2$,
$f_2(T)=\beta_1T^2$ (ii) $f_1(T)=-\Lambda$,
$f_2(T)=\alpha_2T+\beta_2T^2$. We have set the power law cosmology
and developed some constraints on coupling parameter $\lambda$ and
model parameters $\alpha_1$, $\alpha_2$, $\beta_1$ and $\beta_2$. We
have also analyzed the role of these model parameters graphically in
Figures \textbf{(1)}-\textbf{(10)} by exploring the evolution of
WEC, SEC and DEC for both selected models. For the first model, we
have discussed the WEC, DEC, SEC conditions for different cases of
these parameters with $m>1,~0<\Lambda<1$. It is found that WEC can
be satisfied for $\lambda>0$, where we take $0<\alpha,~\beta<1$.
Further, it can also be satisfied for both negative and positive
$\beta$ values if we fix $0<\lambda$ while $0<\alpha<1$. Also, if we
fix both $\beta$ and $\lambda$ as $0<\beta,~\lambda<1$, then WEC can
be satisfied only for $\alpha>0$. Further, it is found that SEC will
be satisfied for this model if $\alpha<0$ and $\lambda>0$ by fixing
other parameters. It is also noticed that DEC will be satisfied for
this model for $\alpha,~\lambda>0$ with fixed values of other
parameters.

For the second model, the involved parameters are
$m,~\gamma,~\delta$ and $\Lambda$ and we have found the appropriate
ranges of these parameters via these bounds. Firstly, we have fixed
$m>1$ and $\Lambda=1$ with $-1<\gamma,~\delta<0$ and it is found
that WEC may be satisfied only for $\lambda>0$. Further, if we set
$0<\lambda<1,~\delta<0$ then WEC remains valid for
$-10\leq\gamma\leq10$. Likewise, if $0<\lambda<1,~\gamma<0$, then
this will be satisfied for $-10\leq\delta\leq10$. We have found that
DEC can be met for negative values of parameters $\gamma$ and
$\delta$ while the SEC can be satisfied for $\lambda>0$, if
$\gamma=1,~\delta>1$. It would be interesting to check these energy
bounds for other models of $f(T)$ gravity and develop the
constraints on the corresponding parameters. 
\end{document}